\documentclass[10pt]{article}
\usepackage{graphicx}
\usepackage{wrapfig}
\usepackage{fancyhdr}
\usepackage[ansinew]{inputenc}
\usepackage{color}
\usepackage{dsfont}

\usepackage{amsmath, amssymb, amsfonts, amsthm, bm, natbib} 
\usepackage{adjustbox}
\def\ds{\displaystyle}
\graphicspath{{./Pictures/}} 
\begin{document} 

\begin{flushleft}
{\Large \bf A Bayesian hidden Markov mixture model to detect overexpressed chromosome regions.\footnote{Accepted for publication in the Journal of the Royal Statistical Society - Series C (Applied Statistics). \\ See http://onlinelibrary.wiley.com/doi/10.1111/rssc.12178/full}}
\end{flushleft}

\vspace{5pt}

{\flushleft 
{\bf Vin\'{i}cius Diniz Mayrink \, and \, Fl\'{a}vio Bambirra Gon\c{c}alves} \\
Departamento de Estat\'{i}stica, Universidade Federal de Minas Gerais, Brazil.}

\vspace{10pt}

\begin{abstract}
In this study, we propose a hidden Markov mixture model for the analysis of gene expression measurements mapped to chromosome locations. These expression values represent preprocessed light intensities observed in each probe of Affymetrix oligonucleotide arrays. Here, the algorithm BLAT is used to align thousands of probe sequences to each chromosome. The main goal is to identify genome regions associated with high expression values which define clusters composed by consecutive observations. The proposed model assumes a mixture distribution in which one of the components (the one with the highest expected value) is supposed to accommodate the overexpressed clusters. The model takes advantage of the serial structure of the data and uses the distance information between neighbours to infer about the existence of a Markov dependence. This dependence is crucially important in the detection of overexpressed regions. We propose and discuss a Markov chain Monte Carlo algorithm to fit the model. Finally, the proposed methodology is used to analyse five data sets representing three types of cancer (breast, ovarian and brain).
\vspace{3pt}
{\flushleft keywords: gene expression; microarray; Affymetrix; cancer; Gibbs sampling.}
\end{abstract}

\begin{figure}[b]
{ \flushleft \small \hspace{0.3cm}\emph{Address for correspondence:}
Vin\'{i}cius Mayrink, Departamento de Estat\'{i}stica, ICEx, UFMG. Av. Ant\^{o}nio Carlos, 6627, Belo Horizonte, MG, Brazil, 31270-901. E-mail: vdm@est.ufmg.br}
\end{figure}

\section{Introduction}

DNA microarray platforms have been widely used in the past years to simultaneously measure the expression levels of a large number of genes; see \cite{Amaratunga} for a comprehensive coverage of recent advances in microarray data analysis. In particular, the high density Affymetrix GeneChip oligonucleotide array technology \citep{Dalma} is the most popular platform used in biomedical researches focused on expression profiling and DNA analysis, at a genome global level or based on a subset of genes. \cite{Irizarry}, \cite{Carvalho}, \cite{Mayrink}, \cite{Li} and \cite{MayrinkBJPS} are few examples of works dealing with this type of data.

In the present study, we are concerned with the identification of chromosome regions associated with high expression measurements obtained from the aforementioned Affymetrix microarrays. In order to link each expression value to a location in the chromosome, we consider the sequence alignment algorithm BLAT \citep{kent} designed to map millions of sequence reads against the human genome. The main advantage of this algorithm when compared to others is its considerably faster performance, which allows for a more regular update of the genome. We provide further details about BLAT in Appendix A.

The identification of differentially expressed probe sets has been the topic of some recent works. In particular, \cite{Warren} proposed a method to generate detection calls ``present", ``marginal" and ``absent" indicating the activity status of genes in the samples. In brief, the authors identify Affymetrix probe sets which cannot hybridize to the intended transcript, and use the empirical cumulative distribution of their intensities to derive a cutoff intensity. A probe set is thus classified as ``Present" if its expression value is higher than the established threshold. The main critique to this method is the arbitrary choice of the cutoff point, which directly affects the classification of probe sets with intensities close to the unknown border separating the high and low categories.

In the present work, we are interested in the identification of overexpressed chromosome regions, based on the BLAT probe expression mapping. In particular, we analyse data from three types of cancer -- breast, ovarian and brain. We consider three data sets from the first type and one data set from each of the other two types. An overexpressed status suggests that a gene plays an important role in the progression of the tumor being investigated; the region identification provided in our study calls the attention for specific parts of the genome where one can find several genes (not only those from the arrays) potentially having a key contribution for understanding the disease.

In order to avoid choosing an arbitrary threshold to classify a value as overexpressed, we take advantage of the irregularly spaced serial structure of the data by using the neighbourhood and distance information. More specifically, we propose a semiparametric hidden Markov model to fit the observations using a mixture distribution with some gammas and a single Gaussian components. The latter is supposed to accommodate the overexpressed clusters. The gamma components are inserted to deal with the skewness and multimodality exhibited by the data. Finally, we consider a mixture of discrete distributions to model the uncertainty on the Markov dependence.

We develop a Markov chain Monte Carlo (MCMC) algorithm to perform inference under the Bayesian paradigm. The algorithm is a Gibbs sampling designed to have good convergence properties. This is achieved by a careful choice of the blocking and update schemes. The most crucial step of the algorithm involves a backward-filtering-forward-sampling strategy.

Mixture models for microarray data have been explored in several previous works. For example, \citet{LBR07} present a Bayesian hierarchical model to determine the expression status of genes using a mixture prior distribution for the parameters representing differential effects. \cite{Broet02} consider a mixture of Gaussians at the data level of the model hierarchy to identify expression changes. A Gaussian mixture is also used in \cite{DoMuller} for the analysis of differentially expressed genes between normal and colon cancer tissue samples. Some other examples are: \cite{Efron01}, \cite{Newton01}, \cite{Parmigiani02}, \cite{BLRDM} and \cite{DR05}. None of these works, however, consider the information regarding the position of the expressions along the chromosomes.

In computational biology, one can easily find studies evaluating data sets containing genome map position information \citep[see][]{Yi,Xu,Cheung,Lucas2010}. In the context of copy number data, \cite{Pollack2002} explores genome-wide measurements of DNA copy number alteration by array CGH; their data are $\log_2$-base fluorescence ratios images depicting amplifications, deletions and unchanged values across the chromosomes. \cite{BR06} propose a novel method called CGHmix to investigate copy number changes, based on a spatially structured mixture model. Other few examples accounting for the spatial dependency along the chromosomes are: \cite{Autio03}, \cite{Jong04} and \cite{Picard05}. Another common research topic using chromosome location data is the study or comparison of genome alignment tools; for example, \cite{Allen} review and compare different approaches to map probes across different microarray platforms.

There is an extensive literature dedicated to methodologies to deal with irregularly spaced serial data. In a continuous-time context, discretely observed diffusion processes might be one option \citep[see][]{Beskos}. Many studies, including the present paper, assume a discrete-time configuration which can be appropriately explored through a particular case of a more general geostatistical model with neighbourhood weights decaying with distance; see, for example, dynamic and state-space models investigated in \cite{Shephard}, \cite{Schnatter} and \cite{dani6}.

One-dimensional change point models may be considered to address the clustering problem accounting for the serial structure of the data. These methods were developed to identify parameter shifts in a distribution assumed for the sequential data. An interesting approach is the Product Partition Model (PPM) introduced by \cite{Hartigan} and \cite{Barry}; its spatial version, presented in \cite{Page}, uses the Euclidean distances between neighbours to determine the probability of the partitions. However,  the model fit would be burdensome to handle thousands of chromosome locations; the PPM generates the unknown change points and this may not be a practical task. As an alternative, \cite{Chib} proposed a model assuming a fixed number of regimes and treating the breaking process as a Markov chain with transition probabilities constrained so that regimes come in a non-reversible sequence. The author indicates that the transition probabilities can be a function of covariates (e.g. the distance information); however, the Markov dependence is assumed true across the whole series despite the distance between neighbours. 

The outline of this paper is as follows: in Section 2, we present all aspects considered to obtain the final data set for analysis; they include: short description of the microarray structure and a scale transformation to better manage the distances between observations. In Section 3, we propose a hidden Markov mixture model to identify overexpressed regions of the chromosomes. Section 4 presents and details the Bayesian inference procedure, including an MCMC algorithm and an strategy to perform the cluster detection based on the output of this algorithm. Section 5 presents the analysis of five data sets representing three types of cancer. Finally, in Section 6, we summarise and discuss the main conclusions of the study.

\section{The data} \label{secdata}

We consider Affymetrix HG-U133A oligonucleotide arrays previously explored in \cite{Miller}, \cite{Wang}, \cite{Sotiriou}, \cite{Marks} and \cite{Freije}. The first three studies are related to breast cancer; the last two refer to ovarian and brain cancer, respectively. Hereafter, we will denote these data sets by ``Breast 1" (251 samples), ``Breast 2" (286 samples), ``Breast 3" (189 samples), ``Ovarian" (141 samples) and ``Brain" (59 samples); these arrays contain expressions of thousands of genes. In brief, an Affymetrix GeneChip consists of a quartz wafer (chip) to which are attached approximately $500{,}000$ different known 25-mer oligonucleotides. The data is obtained through an hybridization procedure where: mRNA are extracted from the cells or tissues of interest, labeled with fluorescent tags and combined with the chip. The mRNA single strands are expected to connect to its complementary sequences, if found, in a specific spot within the array. These array spots are called ``probes'' and they represent a fraction of a gene. Each transcript is represented on an array by a series of 11-20 probe pairs known as a probe set; for simplicity, we may refer to a probe set as a gene. Each pair consists of a perfect match probe (PM), with its 25-base sequence identical to the gene of interest, and a mismatch probe (MM), whose sequence is the same as the PM except for the $13^{\tiny \mbox{th}}$ position, where the base is set to the PM complementary. The MM probe was introduced by Affymetrix as a measure of non-specific binding or cross-hybridization. After the hybridization, the array is washed to remove unconnected material and a laser is applied to activate the fluorescence. Finally, a scanner is used to measure a positive and continuous value of light intensity representing the expression signature in each probe.

In the proposed application, these expression data are mapped to chromosome locations through the alignment tool BLAT -- see Appendix A. Another step to be considered for the raw Affymetrix expression data is a preprocessing routine to remove part of the systematic noise effect aggregated during the experiment to build the arrays. We consider the most popular method known as Robust Multi-chip Average (RMA) \citep[see][]{IrizarryRMA}, also described in Appendix A.
\begin{figure}[!h]
\centering
$$
 \begin{array}{cc}
  \hspace{-0.2cm} \mbox{\tiny (a)} & \hspace{-0.2cm} \mbox{\tiny (b)} \\
  \hspace{-0.2cm} \includegraphics[scale=0.22]{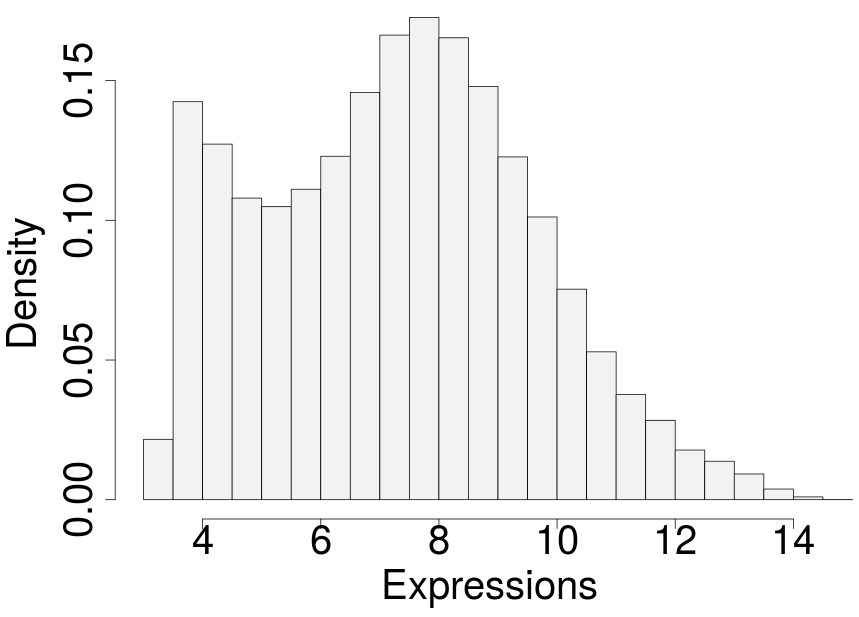} &
  \hspace{-0.5cm} \includegraphics[scale=0.22]{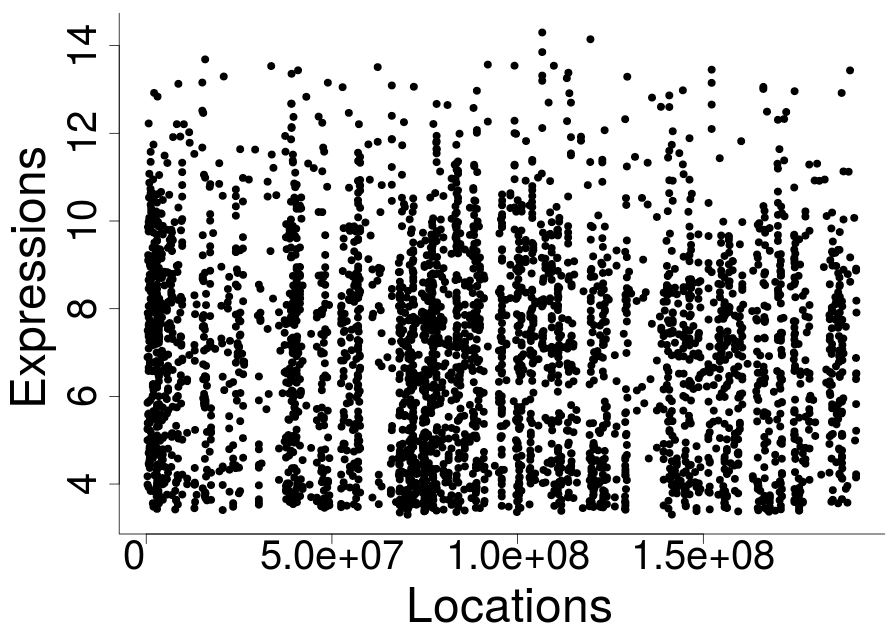} \\
 \end{array}
$$
\vspace{-20pt}
\caption{\scriptsize Graphs related to the data set ``Breast 1". Panel (a): Histogram displaying the distribution of the medians of the preprocessed $\log_2$-base intensities identified via BLAT in all chromosomes (1-22, X, Y). Panel (b): points showing the positions of the medians along chromosome 4.}
\label{MillerChr4}
\end{figure}

Figure \ref{MillerChr4} presents a histogram in panel (a) indicating a skewed and multimodal distribution for the expression data (Breast 1) mapped to the chromosomes. Panel (b) shows the spatial configuration of the data along the chromosome 4 with distances varying between data points. Here, we consider the original scale of chromosome positions identified via BLAT; the minimum, average and maximum distances are: $1$, $53{,}300$ and $3{,}160{,}573$, respectively. The total number of probes mapped to chromosome $4$ is $3{,}582$. Other chromosomes in the human genome have similar results, but their lengths differ; in particular, $Y$ is the shortest one.

The analysis of Figure \ref{MillerChr4} (b) motivates the use of a hypothesis test to verify the spatial dependence in the data. A well known test for this purpose is based on the Moran's I statistics \citep{Moran}. We consider the \texttt{R} package \texttt{spdep} \citep{Bivand} to apply a Monte Carlo version of this test. In brief, the Moran's I, with weights based on the inverse distance between locations, is calculated for the observed data and for many random permutations of the expression values. The resulting p-value represents how often the observed Moran's I is close to those obtained for the permuted data with no spatial association.
\begin{figure}[!h]
\centering
$$
 \begin{array}{cc}
  \hspace{-0.2cm} \mbox{\tiny (a)} & \hspace{-0.2cm} \mbox{\tiny (b)} \\
  \hspace{-0.2cm} \includegraphics[scale=0.22]{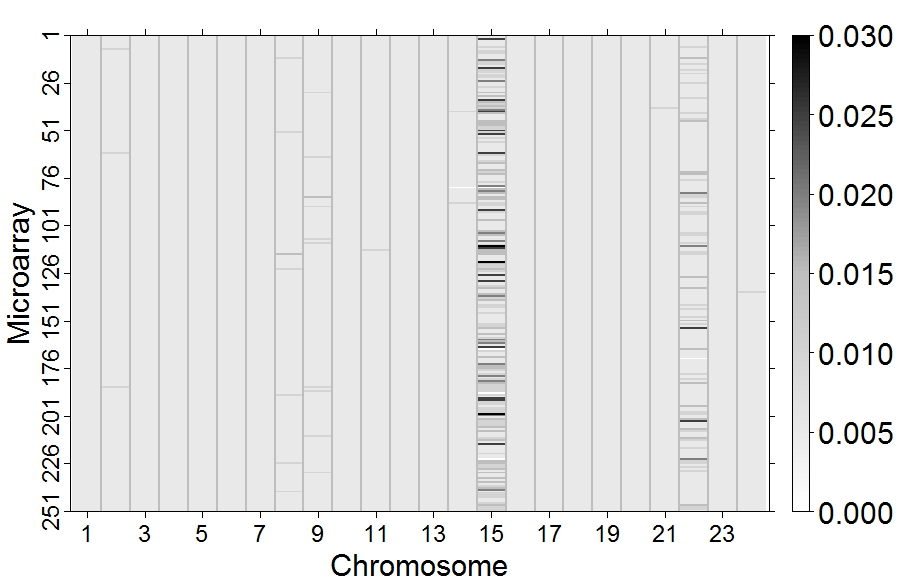} &
  \hspace{-0.2cm} \includegraphics[scale=0.22]{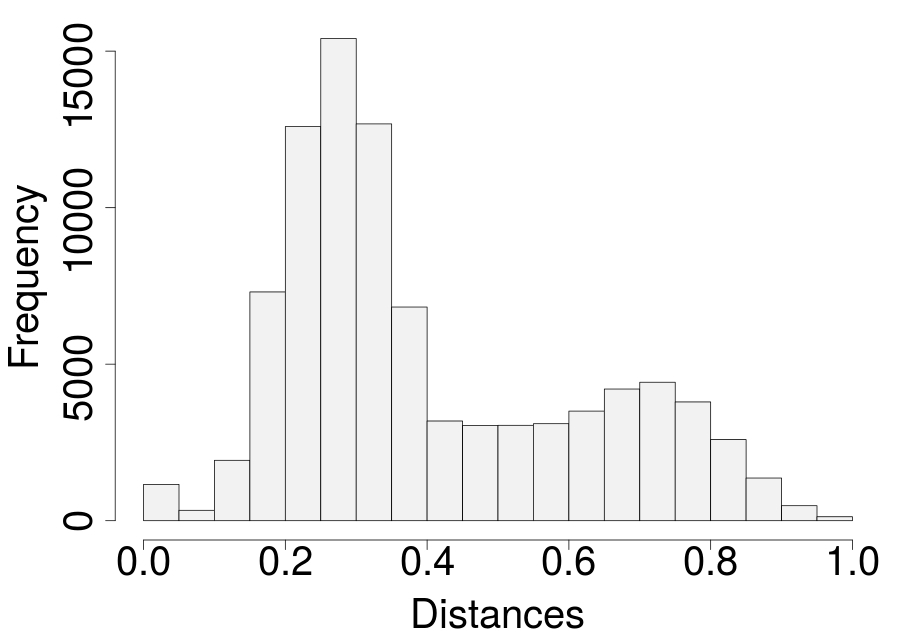} \\
 \end{array}
$$
\vspace{-20pt}
\caption{\scriptsize Panel (a): Image displaying all p-values of the Moran's I permutation test applied to the data set ``Breast 1"; each row represents a microarray and each column represents a chromosome (1-22, X and Y). Panel (b): Histogram of all rescaled distances between locations in all chromosomes.}
\label{MoranDist}
\end{figure}

Figure \ref{MoranDist} (a) shows a heat map graph representing the magnitude of the Moran's I permutation test p-values for each chromosome and each microarray of the ``Breast 1" data; similar results were obtained for the other data sets. Note that all p-values are small ($< 0.05$) suggesting that the spatial association is significant. In particular, chromosome 15 concentrates the largest p-values, but none of them are high enough to reject the association. In summary, these tests indicate presence of spatial dependence in the aligned data and thus motivate a model taking into account the distances between neighbours.

Note that the original BLAT distances are too large and might lead to computational difficulties to fit a model. In order to avoid this issue, we rescale those distances to the interval (0,1) using two steps: calculate the log-distances and divide each of them by the maximum log-distance. The model proposed in the next section will assume the distances as covariates, therefore, other transformations can be considered and this might affect the inference, but it does not change the structure of our model. Taking the logarithm will preserve high differences among the smallest values and practically even out the highest values. This is coherent with the model we propose to perform the analysis, as it will be discussed in the next section. Figure \ref{MoranDist} (b) shows a histogram for all rescaled distances in the chromosomes.

\section{Hidden Markov mixture model} \label{secmodel}

Let $n$ be the number of locations to be analysed and $L$ the number of arrays. Define $X=\{X_i,\;i=1,\ldots,n\}$, $Y=\{Y_i,\;i=1,\ldots,n\}$ and $Y_i=\{Y_{il},\;l=1,\ldots,L\}$, where $X_i$ is the true expression (latent) of location $i$ and $Y_{il}$ is the preprocessed expression of location $i$ observed with noise (w.r.t. $X_i$) in array $l$.

Since the distribution of each $Y_i$ is unimodal, fairly symmetric and $L$ is large (59-286 in our applications), we assume $X_i=\tilde{Y}_i$, for simplicity, where $\tilde{Y}_i$ is the median of $Y_i$. Nevertheless, the variability of the replications could be modeled by assuming, for example, $Y_{il} \sim N(X_i,\tau_{i}^2)$.

The main goal of this analysis is to detect overexpressed regions in each chromosome. The naive attempt of fixing an arbitrary threshold value and looking at the probes with expression intensity higher than this value is discarded for several reasons. Firstly, there is no explicit gap in the distribution of the expression intensity values in each chromosome. Secondly, there is no clear scale of the measured intensity, due to all the preprocessing of the data. Finally, this attempt would not take into account the dependence known to exist among near probes. It is natural to assume that near probes, in the chromosome sequence, are likely to have similar expression intensities. As an example, see the identification of chromosomal regions with DNA copy number alteration in \cite{Pollack2002} and \cite{Lucas2010}.

Our approach considers an stochastic model that uses the dependence among near probes to detect the overexpressed regions. More specifically, the regions of interest are linked to the component of highest mean in a mixture distribution. This way, no fixed threshold needs to be provided. The dependency information is used in a way that near probes are likely to be in the same component of the mixture. Generally speaking, the overexpressed regions are detected as sequences of probes that have the highest expression intensities in the chromosome. The dependence structure disfavors the detection of spikes, i.e. an isolated probe and none of its neighbours being detected as overexpressed. Finally, informative prior distributions for the parameters of the proposed model are crucial to achieve the desired goal.

The multimodality in Figure \ref{MillerChr4} (a) motivates the use of a mixture distribution to model the expressions $X$. This approach also allows the definition of the overexpressed regions, in a stochastic way, by associating them to one of the components in the mixture. This is the component with the highest mean and it is assumed to be a normal distribution. The remaining components of the mixture are assumed to be gamma distributions, which is motivated by the positive, multimodal and skewed features of the data in Figure \ref{MillerChr4} (a) \citep[see][]{Wiper}. One might consider here a mixture of normal distributions instead; however, this would require a larger number of components to fit the data \citep[see][]{Baudry}. Mixtures of gammas have been explored in different contexts \citep[see, for example,][]{Dey,Nascimento}.

Given that BLAT maps thousands of probe expressions to each chromosome and that the majority of locations have neighbours relatively close, it is unlikely that a single location with an intensity much higher than its neighbours is in fact an observation within an overexpressed region. In other words, the neighbourhood structure is crucial in the detection of such regions. This spiked configuration suggests a local atypical expression and is expected to be modelled (associated) to the right tail of a gamma component in the mixture. The local outlier may be explained by cross-hybridization, incorrect probe mapping and other aspects involving the microarray assembling and the alignment tool. The cases where an isolated high expression probe may belong to a potential overexpressed region are those where its nearest observed neighbours are too far away.

We use a Markov structure to model the dependence among probes. However, given the irregular structure of distances between locations in the data, a discrete time Markov chain on (a discretised version of) the distance scale would be both algebraically and computationally expensive. Therefore, we define a Markov structure on the ordering of the probes, i.e. the Markov dependence (if existent) is the same for any consecutive pair of probes. However, such dependence is assumed to be uncertain, i.e. to exist with some probability. In a lower level of the model, we use the distance information to model this probability.

Let $K$ be the number of gamma components in the mixture. This means that the full mixture has $K+1$ components where the one with the largest mean is normal. Define $q_0 = (q_{01},\ldots,q_{0(K+1)})'$ as a probability vector and $Q = \{q_{k_1k_2}\}$, $k_1,k_2 = 1,\ldots,K+1$, as a transition matrix of a $K+1$ states and discrete time Markov chain. Let also $q_k$, $k = 1,\ldots, K+1$, be the $k$-th row of $Q$. Let $F_k$ be the c.d.f. of a gamma distribution with mean $\theta_k$ and shape parameter $\eta_k$, for $k=1,\ldots,K$, and $F_{K+1}$ is the c.d.f. of a normal distribution with mean $\mu$ and variance $\sigma^2$. Denote $f_k$ as the density implied by $F_k$, $k=1,\ldots,K+1$. Define the vector $Z_i = (Z_{i,1},\ldots,Z_{i,K+1})'$ such that $Z_{i,k} = 1$ indicates that $X_i$ belongs to the $k$-th mixture component, 0 otherwise. We propose the following model:
\begin{eqnarray}
  (X_i|Z_{i,k}=1) &\sim& F_k,\;\;i=1,\ldots,n,\;\mbox{all independent}; \label{meq1}\\
  (Z_1|q_0) & \sim & \mbox{Mult}(1,q_0);  \label{meq2} \\
  (Z_i|Z_{i-1,k}=1,\rho_i,q_0,Q) & \sim & (1-\rho_i) \; \mbox{Mult}(1,q_0) + \rho_i \; \mbox{Mult}(1,q_k),\;\;i=2,\ldots,n, \label{meq3}
\end{eqnarray}
where Mult refers to the multinomial distribution and the parameters indexing the mixture components are supressed from the conditional notation. The distribution in (\ref{meq3}) states that a Markov dependence between the expressions in the locations $i-1$ and $i$ is present with probability $\rho_i$.

The model in (\ref{meq1})-(\ref{meq3}) is not an ordinary mixture model because the $X_i$ variables are not marginally (w.r.t. $Z$) independent. Their dependence, however, is defined in a second (and latent) level of the model through a Markov structure. This fact qualifies our model as a Hidden Markov Model (HMM).
For modelling and computational reasons, we define Bernoulli random variables $W_i$, $i=2,\ldots,n$, that indicate the presence of the Markov dependence. That is,
\begin{eqnarray}
  (Z_i|Z_{i-1,k}=1,W_i=0,q_0) & \sim & \mbox{Mult}(1,q_0),\;\;i=2,\ldots,n; \label{meq4} \\
  (Z_i|Z_{i-1,k}=1,W_i=1,Q) & \sim & \mbox{Mult}(1,q_k),\;\;i=2,\ldots,n; \label{meq5} \\
  (W_i|\rho_i) & \sim & \mbox{Ber}(\rho_i),\;\;i=2,\ldots,n. \label{meq6}
\end{eqnarray}
For notation reasons, define $W_1 = 0$ almost surely and $(Z_1|Z_0,W_1=0,q_0):=(Z_1|q_0)$.

Our Bayesian model is fully specified by adopting appropriate prior distributions. Let $\Phi$ be the c.d.f. of the standard normal distribution and define $d_i$ as the (transformed) distance between locations $i-1$ and $i$. We adopt the following prior specifications:
\begin{eqnarray}
  q_0 & \sim & \mbox{Dir}(r_0); \nonumber \\
  q_k & \sim & \mbox{Dir}(r_k),\; \mbox{all independent}; \nonumber \\
  \theta_k & \sim & IG(t_{1k},t_{2k}),\; \theta_1<\ldots<\theta_K; \nonumber \\
  \eta_k & \sim & G(e_{1k},e_{2k}),\;\mbox{all independent}; \label{PriorDist} \\
  (\mu,\sigma^2) & \sim & NIG(m,v,s_1,s_2),\;\mu>\theta_K; \nonumber \\
  (\rho_i|\beta) & = & \Phi(\beta_0+\beta_1d_i); \nonumber\\
  \beta = (\beta_0,\beta_1)' &\sim & N_2(\mu_0,\Sigma_0). \nonumber
\end{eqnarray}
In the specifications above, Dir, $IG$, $G$, $NIG$ and $N_2$ refer to the Dirichlet, inverse gamma, gamma, normal-inverse-gamma and bivariate normal distributions, respectively. Assume the following parametrisation: shape $t_{1k}$ and scale $t_{2k}$ in the inverse gamma prior for $\theta_k$, and mean $e_{1k}$ and shape $e_{2k}$ in the gamma prior for $\eta_k$. Additional notation: $Z = \{Z_i,\;i=1,\ldots,n\}$, $W = \{W_i,\;i=1,\ldots,n\}$,  $\theta = \{\theta_k,\;k=1,\ldots,K\}$ and $\eta = \{\eta_k,\;k=1,\ldots,K\}$.

The Markov dependence of the probes is expressed in terms of the mixture component indicator variable $Z_i$. The existence of this dependence is stochastically explained by the distance between the probes through the probit regression for the $W_i$ variables. Vector $q_0$ could be defined as the stationary distribution of the Markov chain with transition matrix $Q$. However, we do not make this restriction in the model as it would make the inference methodology much harder (more specifically, the sampling step of the $q_k$'s).

The normal component of the mixture is expected to accommodate the expressions of locations that are likely to form the overexpressed regions we are interested in detecting. The symmetry of this distribution is believed to help in the detection. Furthermore, since the clusters of interest are definitely a minority of the probes, we can use this information to elicit informative prior distributions. For example, a prior that concentrates the probabilities $\{q_{0,K+1},q_{1,K+1},\ldots,q_{K,K+1}\}$ around small values. The detection procedure should be performed based on posterior statistics of the $Z_i$'s. We propose a possible approach in Section \ref{clstdet}.

\section{Bayesian inference} \label{secbayes}

Under the Bayesian paradigm, our aim is to obtain the posterior distribution of all unknown quantities involved in the model defined in the previous section. Given the high dimensionality and complexity of this posterior, we devise an MCMC algorithm to sample from it and then use Monte Carlo methods to perform estimation based on this sample.

\subsection{The MCMC algorithm} \label{secmcmc}

The proposed MCMC is a Gibbs sampling with some Metropolis-Hastings (MH) steps. The blocking and sampling schemes are chosen in a way to favor fast convergence of the chain. In this direction, we introduce a set of independent auxiliary variables $V := \{ V_i,\;i=2,\ldots,n \}$ which allows direct sampling from the full conditional distribution of $\beta$ \citep[see][]{albert}. Let $\vec{d}_i = (1, d_i)'$, we set
\begin{equation} \label{auxvrb}
\ds (V_i|\beta) \sim N(\beta' \vec{d}_i, 1)\;\;\;\mbox{and}\;\;\;
    W_i=\left\{ \begin{array}{ll}
                 1, & \mbox{if }V_i>0 \\
                 0, & \mbox{if }V_i\leq0.
            \end{array} \right. \nonumber
\end{equation}
Note that the original model for $X$ is preserved -- simply integrate $V$ out to check it.

Let $\psi := \{ \theta, \mu, \sigma^2 \}$, we choose the following blocking scheme for the Gibbs sampler:
\begin{equation}\label{bloc}
\ds (Z,W)\;\;;\;\;V\;\;;\;\;(q_0,Q,\beta,\psi)\;\;;\;\;\eta. \nonumber
\end{equation}
This scheme leads to a non-irreducible Markov chain which, in turn, does not converge. We circumvent this problem by making the algorithm a collapsed Gibbs sampling where $V$ is integrated out from the full conditional distribution of $(Z,W)$. This strategy makes the chain irreducible and guarantees its convergence to the target distribution \citep[see][]{CGS}.

All the full conditional densities from the Gibbs sampler are proportional to the joint density of $X$ and all the unknown quantities of the model -- this density is given in (\ref{jdens}), in Appendix B. We now describe each step of the algorithm.

\vspace{5pt}

{\bf Sampling $V$} \vspace{3pt}

The full conditional distribution of $V$ is such that the $V_i$'s are all independent with distribution $\ds N(\beta'\vec{d}_i, 1)$, truncated to be positive if $W_i=1$ and non-positive if $W_i=0$.

\vspace{5pt}

{\bf Sampling $(q_0, Q, \beta, \psi)$} \vspace{3pt}

The four components of this block -- $q_0$, $Q$, $\beta$ and $\psi$, are conditionally independent. This means that a draw from this block is obtained by sampling each of the four components individually from their respective marginal full conditional distribution. The marginals of the first three components are given in (\ref{fc1})-(\ref{fc3}), in Appendix B.

The marginal full conditional distribution of $\psi$ is a truncation of a tractable and easy to simulate distribution, which is defined in (\ref{fc4}). The truncation region is defined by the restriction $\{\theta_1 < \theta_2 < \ldots <\theta_K < \mu\}$. We sample exactly from this truncated distribution via rejection sampling by proposing from its non-truncated version and accepting if the restriction is satisfied. Since $K$ is chosen to be small, this algorithm is computationally efficient.

\vspace{5pt}

{\bf Sampling $\eta$} \vspace{3pt}

The $\eta_k$'s are sampled (jointly or individually) via Gaussian random walk MH step(s) properly tuned to have reasonable acceptance rates \citep[see][]{rgg}. Details on the proposal distribution and acceptance probability of the MH step where each $\eta_k$ is sampled separately are given in (\ref{MH1}) and (\ref{MH2}), respectively, in Appendix B.

\vspace{5pt}

{\bf Sampling $(Z,W)$} \vspace{3pt}

This is the most challenging step of the MCMC. Although the full conditional can be factorised into multinomial distributions, obtaining such factorization and the parameters of each multinomial is not straightforward. Firstly, note that integrating out $V$ provides the following full conditional kernel:
\begin{eqnarray}\label{fcZW}
\ds \pi(Z,W|\cdot) & \propto & \prod_{i=1}^n\left[\prod_{k=1}^{K+1}\left[ (f_k(X_i|\psi)q_{0k})^{Z_{ik}}\right]^{1-W_i} \left[(f_k(X_i|\psi)q_{k_{(i-1)}k})^{Z_{ik}}\right]^{W_i}\right] \nonumber \\
& & \times \; \left( \Phi_i^+ \right)^{W_i}\left( \Phi_i^- \right)^{1-W_i}\prod_{k=1}^{K+1}q_{0k}^{Z_{0k}},
\end{eqnarray}
with $\Phi_i^+ = \Phi(\beta'\vec{d}_i)$ and $\Phi_i^- = \Phi(-\beta'\vec{d}_i)$.

Direct sampling from (\ref{fcZW}) is only possible if we adopt the following factorization:
\begin{equation}\label{fcZ2}
\ds \pi(Z,W|\cdot)\propto\pi(Z_1|\cdot)\prod_{i=2}^n\pi(Z_i|W_i,Z_{i-1},\cdot)\pi(W_i|Z_{i-1},\cdot), \nonumber
\end{equation}
which means that $(Z,W)$ is sampled in the following order: $Z_1,W_2,Z_2,\ldots,W_n,Z_n$, i.e. forward in time. Each marginal distribution is multinomial (Bernoulli for the $W_i$'s) for which the parameter values are obtained recursively and backward in time. Therefore, we denote this sampling scheme as a backward-filtering-forward-sampling (BFFS). Details on this sampling step are given in Appendix B.

The MCMC algorithm described here was implemented using the \texttt{R} programming language \citep{softwareR}.

\subsection{Cluster detection}\label{clstdet}

As it was mentioned before in Section \ref{secmodel}, the clusters of interest are expected to be accommodated by the Gaussian component of the mixture. In this sense, the cluster detection should be performed based on the posterior probability of one or more probes belonging to the Gaussian component of the mixture.
Given a (approximate) sample of size $M$ from the posterior distribution of $Z$ and a sequence $\{i_1,i_2,\ldots,i_S\}$ of expressions ($S = 1, 2, \ldots$), the posterior probability of this sequence being a cluster of interest is given by:
\begin{equation}\label{clustprob}
\ds \frac{1}{M}\sum_{m=1}^M\prod_{s=1}^S \mathds{1}(Z_{i_s,K+1}^{(m)}=1),
\end{equation}
where $\mathds{1}(.)$ is an indicator function.

A practical (computationally efficient) procedure for cluster detection is to evaluate the probability in (\ref{clustprob}) for each individual probe and look for sequences of probes of a minimum size (say 4 or 5) for which the individual probabilities are high (say $> 0.5$). For such sequences, the posterior probability in (\ref{clustprob}) will be considerably high (given the Markov dependence imposed by the model). One may also be interested in evaluating this probability for particular sequences -- given some practical reason. In Section \ref{secresult}, we search for a sequence (size 5) of high individual posterior probabilities and report its corresponding joint posterior probability for each cancer data set.

\section{Results for the cancer data sets} \label{secresult}

We consider five different sets of microarray data representing three types of cancer (see Section \ref{secdata}); the number of samples varies from 59 to 286 arrays. Since the expression scale is the same for every chromosome, we perform a joint analysis of all the 24 chromosomes (for each of the five data sets). Moreover, given that there is no reason to believe that there exists a dependence between the last and first probes of consecutive chromosomes (there is no distance to be measured), we fix the $W_i$ corresponding to the first probe of each chromosome to be 0. An important implication of this is that the BFFS step to sample $(Z,W)$ may be performed separately for each chromosome, given its conditional independence property. This leads to a significant improvement in the computational cost.

An important decision to fit the model is to choose the number $K$ of gamma components in the mixture distribution. The histogram in Figure \ref{MillerChr4} (a) clearly indicates that a single gamma distribution would not be appropriate for this analysis due to the multimodal and asymmetric shape of the graph. In order to investigate other reasonable possibilities, we fit the model with $K =$ 2, 3, 4 and 5 to the ``Breast 1" data set (the distributions of the expressions have similar behavior for all data sets). Results are reported in Appendix D and indicate that $K=4$ is the best choice. The model with $K = 2$ led to a wide (high variance) Gaussian component including too many expressions (some of them very low). Results for $K = 3$ are similar but still significantly different from those with $K=4$ which, in turn, are virtually the same as for $K=5$ (especially in terms of overexpressed region detection).

In terms of prior specification, consider again the notation in (\ref{PriorDist}). We assume $r_0 = (750,750,750,750, 10)'$ as the parameter vector of the Dirichlet prior for $q_0$; here, we indicate that an expression at a location without a Markov dependence most likely belongs to a gamma component in the mixture. The large equal weights for the gammas are required in this application involving thousands of locations; small concentration parameters would lead to a weakly informative prior dominated by the data.
This strategy is also considered for the prior specification of $q_k$. We assume the following matrix:
{\small
$$
r = \left(
\begin{array}{ccccc}
 969.70 & 484.85 &   96.97 &   48.48 &   10.0 \\
 484.85 & 969.70 & 484.85 &   96.97 &   10.0 \\
   96.97 & 484.85 & 969.70 & 484.85 &   10.0 \\
   48.48 &   96.97 & 484.85 & 969.70 &   10.0 \\
   48.48 &   48.48 &   96.97 & 484.85 & 969.7 \\
\end{array}
\right).$$}
The actual values of the hyperparameters for $q_0$ and $Q$ are established in a way that the prior had the same weight for the Gaussian component and the same level of information for each specification of $K$ in the above mentioned sensitivity analysis (rounded numbers were fixed for $K=3$). The $k$-th row of matrix $r$ is the parameter vector of the Dirichlet prior for $q_k$; recall that here we account for the Markov dependence. Note that the largest weights were defined in the main diagonal to favor the model to allocate the $i$-th observation in the same component of its neighbour from location $i-1$. This matrix also has weights decreasing as we move away from the main diagonal elements -- this configuration is used to discourage the model to assign two consecutive observations in distant components. Another important aspect of this prior are the small weights specified for the Gaussian component in the last column of the first four rows and the high weight specified in the last row; this choice is an strategy to make this component accessible only to those observations having great evidence of overexpression and having neighbours with the same characteristics.

The bivariate normal distribution for the coefficients ($\beta_0$, $\beta_1$) is set with mean ($4$,$-8$) and covariance matrix $10 \, \bm{I_2}$; i.e., we assume prior independence for these coefficients and the mean indicates the following information about the probability of having a Markov dependence: $\lim_{d_i \rightarrow 0} \rho_i \approx 1$, $\lim_{d_i \rightarrow 1} \rho_i \approx 0$ and $\rho_i = 0.5$ for $d_i = 0.5$.

In this application, we assume a gamma prior distribution with mean 50 and shape 1 for the shape parameter $\eta_k$ of all the four gamma components in the mixture; its variance is $50^2$. In addition, the mean $\theta_k$ has the following inverse gamma prior specification: $t_{1k} = 4$ (for $k = 1,2,3,4$), $t_{21} = 9$, $t_{22} = 18$, $t_{23} = 27$ and $t_{24} = 36$. These hyperparameter choices provide the expected values ($3$, $6$, $9$, $12$) and the standard deviations ($2.12$, $4.24$, $6.36$, $8.48$) for $k = 1,2,3,4$, respectively. We have chosen the expected values based on the scale of the observations and the role we expect the gamma components to play in the model. Finally, we specify a normal-inverse-gamma prior with parameters ($15$, $25$, $2.1$, $1.1$) for the pair ($\mu$, $\sigma^2$) related to the Gaussian component. This implies that $E(\sigma^2) = 1$ and $\mbox{Var}(\sigma^2) = 10$. A prior sensitivity analysis was performed by doubling the standard deviations of $(\mu, \sigma^2, \theta_1, \ldots, \theta_4)$ and the very same results were obtained -- see Appendix D.

\begin{table}[!h]
\caption[caption]{\label{tabMeanSD} \scriptsize Posterior mean and standard deviation (in parentheses) for the parameters of the mixture distribution with four gamma \\\hspace{\textwidth}
and one Gaussian components.}
{\scriptsize\begin{adjustbox}{center}
\fbox{%
\begin{tabular}{crrrrr}
 \noalign{\smallskip}
            & \em Breast 1 & \em Breast 2 & \em Breast 3  & \em Ovarian & \em Brain  \\
 \hline
 \noalign{\smallskip}
  $\theta_1$   & 4.020 (0.008)     & 3.769 (0.010)     & 4.127 (0.008)     & 3.642 (0.007)     & 4.365 (0.007) \\
  $\eta_1$      & 105.775 (2.920) & 112.609 (4.107) & 120.475 (3.386) & 177.937 (5.546) & 268.408 ( 8.919) \\
  \noalign{\smallskip}
  \noalign{\smallskip}
  $\theta_2$   & 6.398 (0.028)     & 5.789 (0.051)     & 6.351 (0.029)     & 4.991 (0.029)     & 5.612 (0.032) \\
  $\eta_2$      & 20.815 (0.326)   & 22.487 (0.573)   & 22.724 (0.399)   & 36.937 (0.995)   & 52.620 (1.721) \\
  \noalign{\smallskip}
  \noalign{\smallskip}
  $\theta_3$   & 8.066 (0.021)     & 7.684 (0.021)     & 8.080 (0.023)     & 7.213 (0.025)     & 7.776 (0.030) \\
  $\eta_3$      & 62.996 (1.596)   & 65.062 (1.500)   & 57.345 (1.556)   & 51.707 (1.238)   & 61.648 (1.546) \\
  \noalign{\smallskip}
  \noalign{\smallskip}
  $\theta_4$   & 9.824 (0.028)     & 9.473 (0.033)     & 9.882 (0.027)     & 8.846 (0.029)     & 9.573 (0.031) \\
  $\eta_4$      & 94.436 (2.528)   & 87.524 (2.750)   & 93.658 (2.547)   & 68.373 (1.982)   & 88.486 (2.623) \\
  \noalign{\smallskip}
  \noalign{\smallskip}
  $\mu$          & 12.132 (0.033)   & 11.997 (0.036)   & 12.261 (0.032)   & 11.504 (0.034)   & 11.780 (0.034) \\
  $\sigma^2$ & 0.831 (0.034)     & 0.947 (0.040)     & 0.843 (0.033)     & 1.103 (0.039)     & 0.873 (0.033) \\
\end{tabular}}
\end{adjustbox}}
\end{table}

\vspace{-10pt}

\begin{table}[!h]
\caption[caption]{\label{tabWeights} \scriptsize Posterior mean of the coordinates of the $Z_{i}$'s -- posterior weights of the mixture components.}
{\scriptsize\begin{adjustbox}{center}
\fbox{%
\begin{tabular}{crrrrr}
 \noalign{\smallskip}
 $k$ & \em Breast 1 & \em Breast 2 & \em Breast 3 & \em Ovarian & \em Brain \\
 \hline
 \noalign{\smallskip}
 1 & 0.138 & 0.150 & 0.140 & 0.120 & 0.124 \\
  \noalign{\smallskip}
 2 & 0.387 & 0.378 & 0.381 & 0.278 & 0.269 \\
  \noalign{\smallskip}
 3 & 0.274 & 0.286 & 0.280 & 0.371 & 0.365 \\
  \noalign{\smallskip}
 4 & 0.165 & 0.148 & 0.162 & 0.188 & 0.198 \\
  \noalign{\smallskip}
 5 & 0.036 & 0.037 & 0.037 & 0.042 & 0.045 \\
\end{tabular}}
\end{adjustbox}}
\end{table}

\vspace{-10pt}

\begin{table}[!h]
\caption[caption]{\label{Intersect} \scriptsize Comparison involving pairs of data sets. Number of probes identified in the Gaussian component in both data sets \\\hspace{\textwidth}
of the pair, relative to the number of probes identified in the Gaussian component in at least one of the data sets of the pair.}
{\scriptsize\begin{adjustbox}{center}
\fbox{%
\begin{tabular}{crrrrr}
 \noalign{\smallskip}
            & \em Breast 1 & \em Breast 2 & \em Breast 3  & \em Ovarian & \em Brain  \\
 \hline
 \noalign{\smallskip}
 \em Breast 1 & 100\% & 79.1\%  & 87.4\% & 71.6\% & 62.9\% \\
 \noalign{\smallskip}
 \em Breast 2 &            & 100\%   & 84.6\% & 76.1\% & 64.5\% \\
 \noalign{\smallskip}
 \em Breast 3 &            &              & 100\%  & 75.4\% & 65.9\% \\
 \noalign{\smallskip}
 \em Ovarian  &            &              &             & 100\%  & 69.3\% \\
 \noalign{\smallskip}
 \em Brain      &            &              &             &             & 100\%  \\
\end{tabular}}
\end{adjustbox}}
\end{table}

The MCMC is set to perform $15{,}000$ iterations with a burn-in of $5{,}000$. Let $\bm{1_{(l_1 \times l_2)}}$ be a $l_1 \times l_2$ matrix of ones; in terms of initial values, we consider $q_0^{(0)} = 0.2 \, \bm{1_{(5 \times 1)}}$ and $Q^{(0)} = 0.2 \, \bm{1_{(5 \times 5)}}$. In addition, we set $\beta_0^{(0)} = 4$, $\beta_1^{(0)} = -8$ as specified in their prior distribution. The initial values of the remaining parameters are set based on statistics of the data; we break the support of the histogram (see Figure \ref{MillerChr4}) into 4 contiguous intervals and then evaluate their means and variances to determine: $\theta_1^{(0)} = 4.43$, $\theta_2^{(0)} = 6.70$, $\theta_3^{(0)} = 8.96$, $\theta_4^{(0)} = 11.23$, $\mu^{(0)} = 13.50$, $\eta_1^{(0)} = 50.87$, $\eta_2^{(0)} = 108.30$, $\eta_3^{(0)} = 200.30$, $\eta_4^{(0)} = 330.93$ and $(\sigma^2)^{(0)} = 0.21$. The variances of the random walk proposals in the MH step to sample $\eta_1$, $\eta_2$, $\eta_3$ and $\eta_4$ were chosen to provide reasonable (around $44\%$) acceptance rates. Convergence is rapidly attained -- Appendix E shows some diagnostics.

\begin{figure}[!h]
\centering
   \includegraphics[scale=0.32]{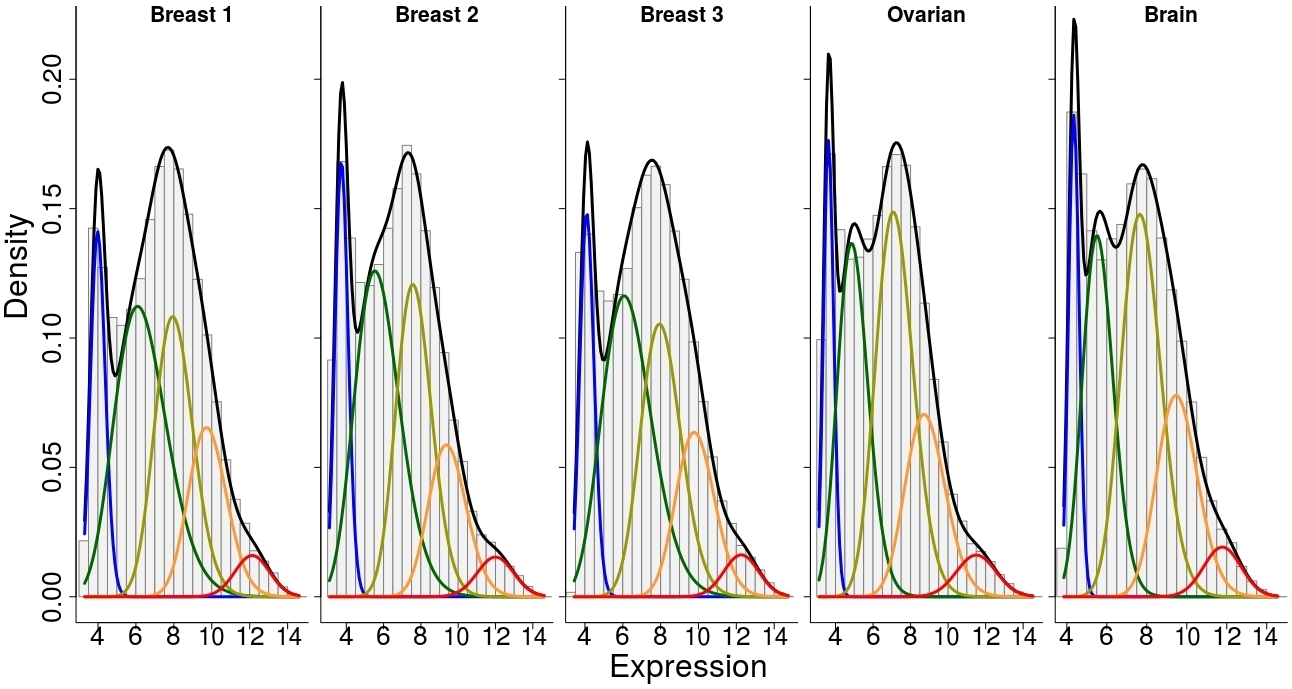}
\vspace{-10pt} \caption{\scriptsize Histogram of all expressions overlaid by the estimated mixture density (black curve) and its components (gammas in blue, green, yellow and orange, normal in red).}
\label{figMixDensity}
\end{figure}

\vspace{-10pt}

\begin{figure}[!h]
\centering
   \includegraphics[scale=0.26]{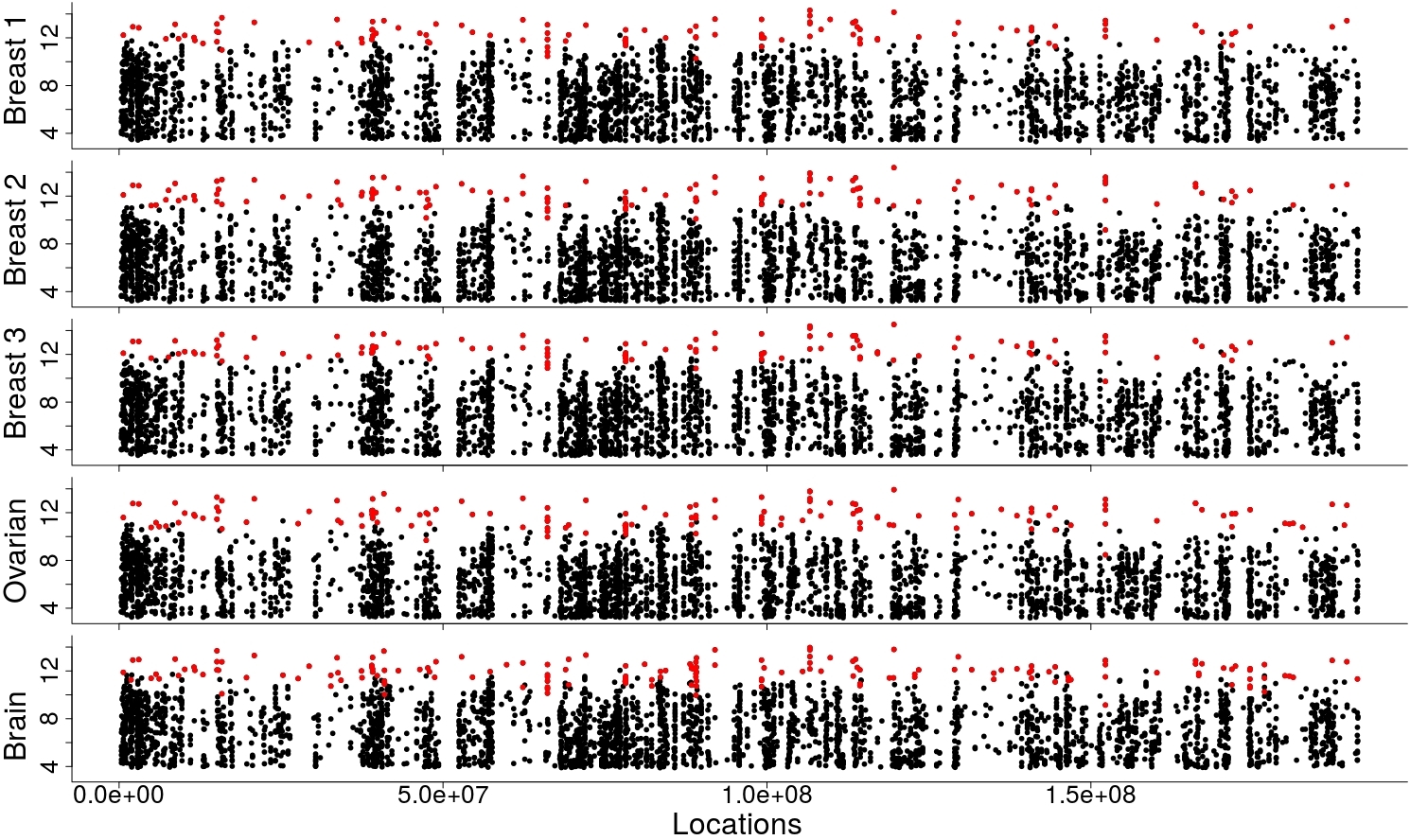}
\vspace{-10pt} \caption{\scriptsize Expression values along chromosome 4. The red dots indicate observations where the Gaussian component has posterior probability above 0.5.}
\label{chr4Points}
\end{figure}

Table \ref{tabMeanSD} presents the posterior estimates for each component of the mixture distribution and Table \ref{tabWeights} their corresponding posterior weight (posterior mean of the coordinates of the $Z_{i}$'s). For each parameter, note that the means and standard deviations are similar across the data sets; this suggests mixture densities having nearly the same shape when comparing different cancers. 

Figure \ref{figMixDensity} shows the histogram of the data overlaid by the estimated mixture density and their components for each data set. Note that the range of the Gaussian is not too wide and, therefore, accommodates the expressions in the right tail of the histograms. This result suggests that the mixture with $K = 4$ gamma components represents well the data.

Figure \ref{chr4Points} shows the behaviour of the expression values along chromosome 4 for all five data sets; here, we explore a single chromosome since the graph displaying all of them would be difficult to visualise. As can be seen, the red dots, representing observations with high posterior probability ($> 0.5$) of belonging to the Gaussian component, are concentrated on the top of the graphs. Note that non-Gaussian observations (black dots) can be identified among Gaussian ones, which indicates high expressions originated from the right tail of a gamma distribution in the mixture. This occurs as a result of the model structure accounting for distances between locations to determine the strength of the Markov dependence. Again, the simple strategy of choosing a threshold above which all points are in red would inflate the number of overexpressed regions.

We also compute the posterior probability of a specific cluster of 5 locations being an overexpressed region for each data set -- see (\ref{clustprob}). We evaluate the sequence composed by locations $33{,}312$ to $33{,}316$. Its posterior probability of being an overexpressed cluster is: $0.9996$ (Breast 1), $0.9998$ (Breast 2), $0.9998$ (Breast 3), $0.9998$ (Ovarian) and $0.9989$ (Brain).

Another interesting aspect displayed in Figure \ref{chr4Points} is the fact that for the same location the expressions from each data set are similar, consequently, the cluster identification tends also to be similar comparing these data sets. If we consider all 24 chromosomes ($91{,}090$ locations identified via BLAT), the number of probes in the Gaussian component are: $3{,}132$ (Breast 1), $3{,}277$ (Breast 2), $3{,}271$ (Breast 3), $3{,}723$ (Ovarian) and $3{,}860$ (Brain). Table \ref{Intersect} compares each pair of data sets and indicates the frequency of locations identified in the Gaussian component in both data sets, relative to the total number of locations identified in the Gaussian component in at least one of them. Note that the largest percentages ($79.1\%$--$87.4\%$) are obtained for pairs involving two breast cancer data sets; this is a plausible result suggesting that the model is working well in this cluster identification problem involving $91{,}090$ locations. The overexpressed regions in the breast cancer data sets seem to have higher similarity with the ovarian cancer ($71.6\%$--$76.1\%$) than with the brain cancer ($62.9\%$--$69.3\%$).

\begin{figure}[!h]
$$
  \begin{array}{cc}
   \hspace{-0.4cm} \mbox{\tiny (a)} & \hspace{-0.5cm} \mbox{\tiny (b)} \\
   \hspace{-0.4cm} \includegraphics[scale=0.21]{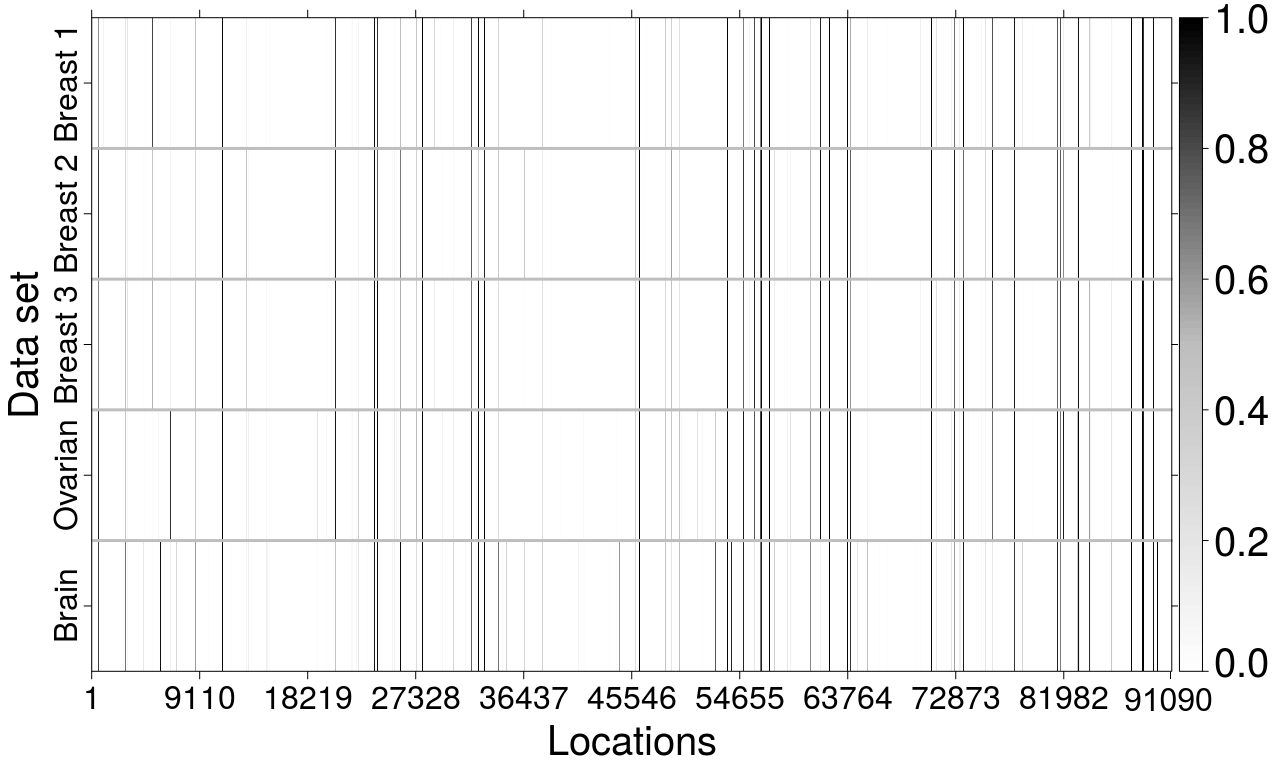} & \hspace{-0.2cm} \includegraphics[scale=0.21]{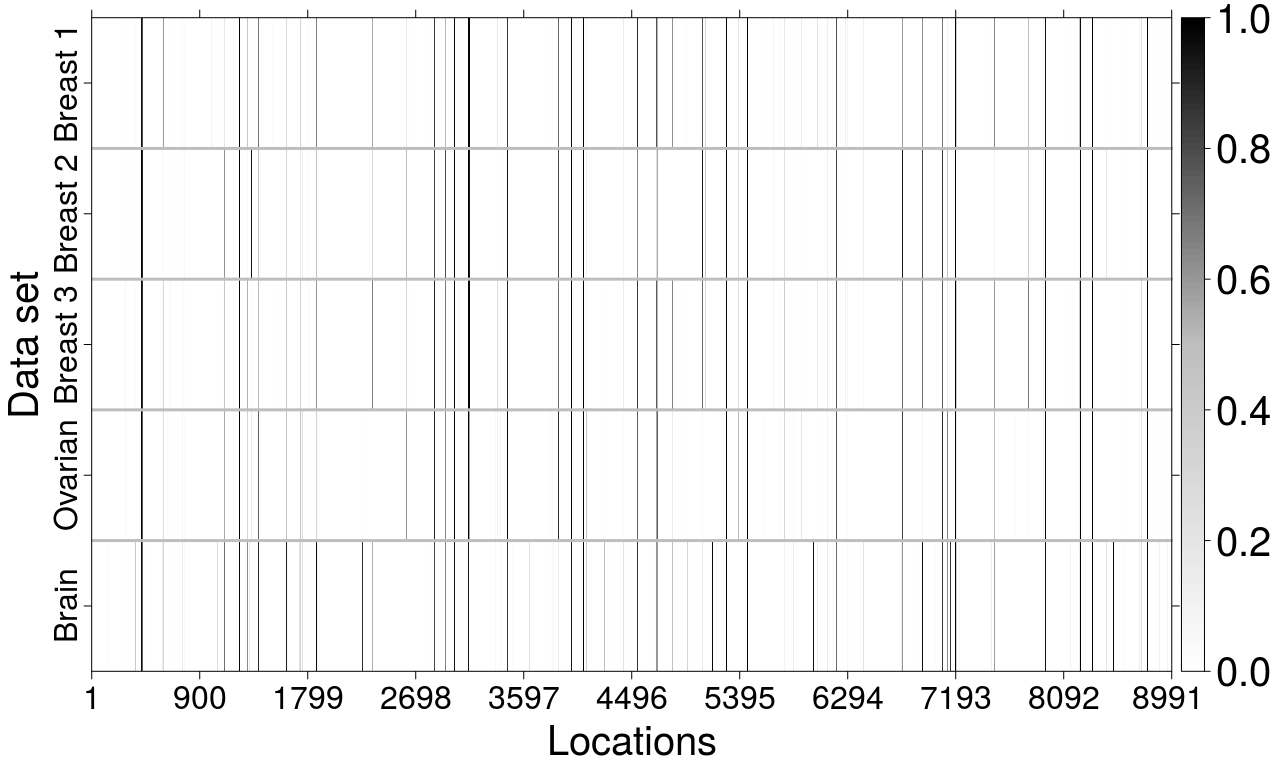}\\
  \end{array}
$$
\vspace{-15pt} \caption{\scriptsize Heat map image indicating for each location the posterior probability of belonging to the Gaussian component. Panel (a) represents all chromosomes and Panel (b) shows chromosome 1.}
\label{HeatMaps}
\end{figure}

\vspace{-10pt}

\begin{figure}[!h]
$$
  \begin{array}{cc}
	 \hspace{-0.2cm} \mbox{\tiny (a)} & \mbox{\tiny (b)} \\
   \hspace{-0.2cm} \includegraphics[scale=0.20]{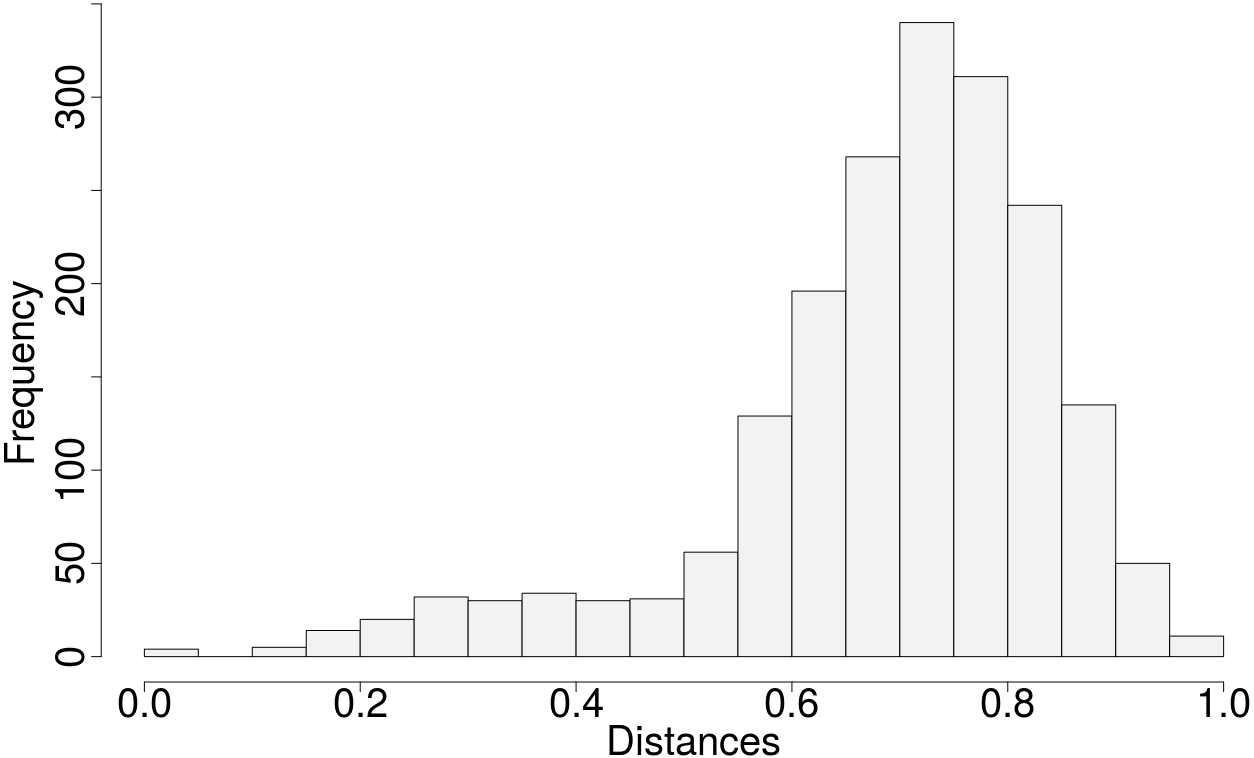} & \hspace{-0.2cm} \includegraphics[scale=0.20]{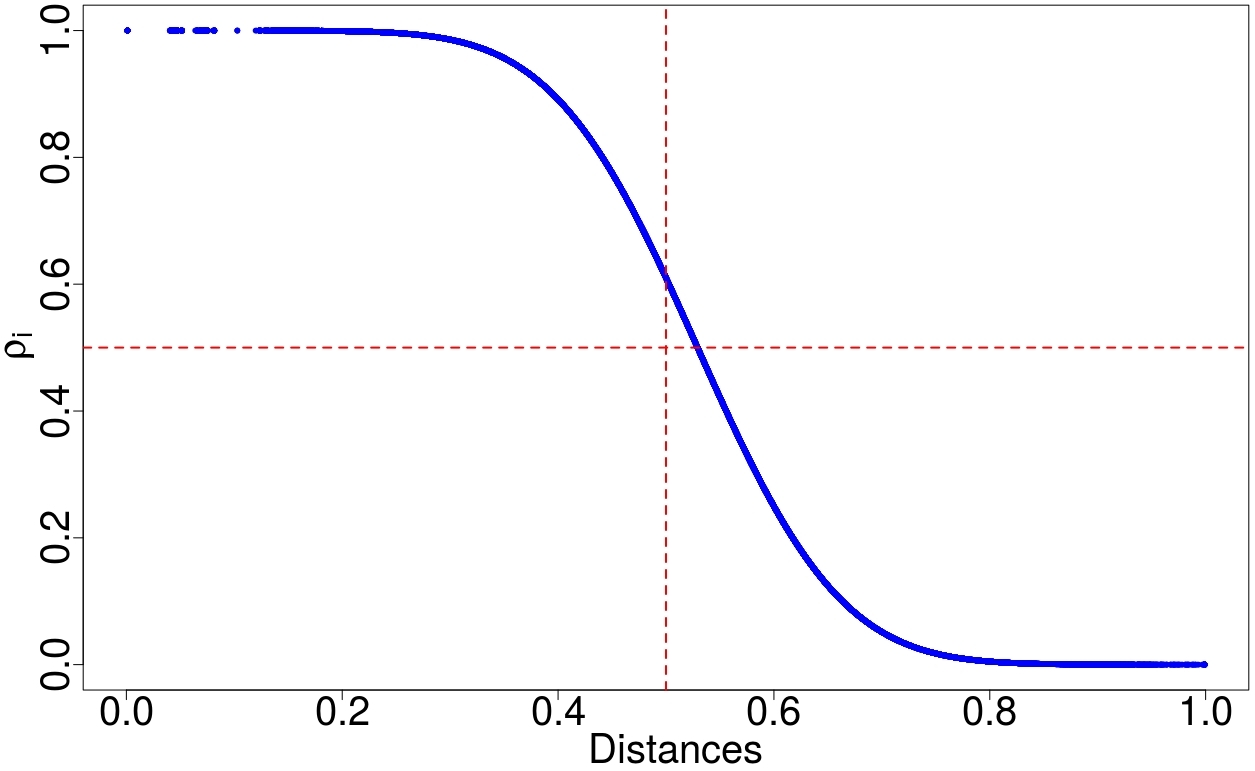} \\
  \end{array}
$$
\vspace{-15pt} \caption{\scriptsize Panel (a): Histogram of the distances between the location of a cluster, with a single observation, and its nearest non-Gaussian neighbours. Panel (b): Estimated relationship between the distances and $\rho_i$ (probability of having a Markov dependence with the previous probe); here, we consider the ``Breast 1" data set and the posterior mean of $\beta_0$ (5.05) and $\beta_1$ (-9.55) to draw the curve.}
\label{NeighRho}
\end{figure}

Figure \ref{HeatMaps} shows a grey scale heat map investigating the posterior probabilities of being a Gaussian component observation. The results for all 24 chromosomes are shown in panel (a) and only the chromosome 1 is explored in panel (b); images for the chromosomes 2-22, $X$ and $Y$ have a similar behaviour and are presented (individually) in Appendix C. The conclusions here reinforce the previous analysis of Figure \ref{chr4Points} and Table \ref{Intersect}; i.e., the data sets tend to agree, exhibiting similar probabilities for the same location. Note that a coherent pattern of grey/black probabilities can be clearly seen in both panels and the majority of locations are in white due to a estimated probability of Gaussian classification near 0 for most chromosomal regions. This coherence between results from different microarray data sets, specially between those three for breast cancer, is a strong evidence that the proposed model is performing well. A disagreement between data sets for the same type of tumor is not expected and, assuming that the data were correctly preprocessed and aligned, this would suggest a critical model issue; this is not the case here.

In this application, we have observed different sizes of sequences of consecutive locations belonging to the Gaussian component when adopting the classification rule described in the caption of Figure \ref{chr4Points}. The shortest case is a single observation surrounded by two non-Gaussian neighbours; the longest sequence, comparing all data sets, involves 17 locations. One may consider that the identification of a single Gaussian location with non-Gaussian neighbours is potentially a model issue; however, the histogram in Figure \ref{NeighRho} (a) shows that this is not the case. It shows that the distribution of all distances between a single Gaussian location and its non-Gaussian neighbours is highly concentrated in the region above 0.5. Therefore, the magnitude of these isolated expressions is compatible with an overexpressed cluster and the model cannot use the neighbourhood information to change this classification because the neighbours are too far away. These locations belong to potential clusters but a confirmation relies on their non-observed neighbours.

Figure \ref{NeighRho} (b) presents the estimated relationship between the distances and the probability $\rho_i$ of having a Markov dependence at location $i$ for the ``Breast 1" data; two dashed lines were included to highlight the probability and the distance 0.5. The posterior means of ($\beta_0$,$\beta_1$) for each data set are: (5.05,-9.55) in ``Breast 1", (5.02,-9.50) in ``Breast 2", (5.26,-9.85) in ``Breast 3", (5.27,-9.83) in ``Ovarian" and (4.92,-9.31) in the ``Brain" data set. These are all similar, which implies that their corresponding curves will be very close to the one in panel (b). We can evaluate the decreasing behaviour of the curve to understand the impact of the distances on $\rho_i$; for example, the slow decay before the distance 0.4 indicates that the Markov dependence is strong for those distances. In addition, the curve has a fast decay between distances 0.4 and 0.6, and is approximately 0 from 0.8 onwards.

\section{Conclusions}

In this paper, we have developed a hidden Markov model designed for an application involving Affymetrix DNA microarray data. The study focuses on the identification of chromosomal regions associated with high gene expression measurements, which we call overexpressed regions. The microarray probe expressions are mapped to locations in the human chromosomes using the alignment algorithm BLAT. As a result, the data is configured as an irregularly spaced sequence along the chromosomes; five different data sets representing breast, ovarian and brain cancers were considered in the analysis. The original light intensities from the arrays were preprocessed via RMA.

The proposed model assumes a mixture distribution with four gamma and one normal components to cluster the observations. The largest mean is imposed for the Gaussian component, which is supposed to accommodate the target overexpressed values. The model takes advantage of the distances between the identified locations to determine whether there is a Markov dependence between neighbours and uses such dependence to (stochastically) define the target clusters. Inference is performed under the Bayesian paradigm via an MCMC algorithm consisting of a carefully devised Gibbs sampling.

The results indicate that the model is selective in the cluster determination, being able to discriminate well the locations to identify the overexpressed regions; on average, only $3.8\%$ of the locations are classified (for a suitable classification rule) in potential overexpressed clusters. Looking at the intersection of overexpressed regions from the analised data sets and evaluating the coherence pattern exhibited in image graphs, our findings suggest great similarity between them. As expected, the greatest similarity is observed among the breast cancer data sets, which is a strong indication that the model is performing well. The real application also shows that the breast cancer data is more similar to the ovarian cancer than to the brain cancer.

We perform a general cluster identification analysis to present a global picture of the overexpressed regions across the human genome. This means that the results obtained here can be used for a more detailed description of specific regions of interest in any chromosome; for example, one can easily compute the probability of a particular set of probes being an overexpressed cluster through the cluster detection procedure presented in (\ref{clustprob}).

\vspace{15pt}
{\flushleft \sffamily \textbf{Acknowledgements}}
\vspace{5pt}

The authors would like to thank the Associate Editor and the anonymous referee for their constructive comments leading to a much improved version of the paper. The authors also thank FAPEMIG for supporting this research. The first author would like to thank Joseph E. Lucas for the initial discussion about this topic at Duke University.

\vspace{5pt}
\renewcommand{\thefigure}{A.\arabic{figure}} \setcounter{figure}{0}
\renewcommand{\theequation}{A.\arabic{equation}} \setcounter{equation}{0}
\renewcommand{\thetable}{A.\arabic{table}} \setcounter{table}{0}
{\flushleft \textbf{Appendix A: Details about BLAT and RMA} }
\vspace{10pt}

The BLAT algorithm was developed to quickly find genome positions having at least 95\% similarity with sequences of interest in a database. In fact, BLAT is available for different types of query sequences and, in our application, we use this algorithm on DNA to search for chromosome locations having high compatibility with the 25-bases probe sequences from the microarray. The shorter is the length of the sequence, the higher is the risk of incorrect mapping; this issue is not peculiar to BLAT and it can also occur in other alignment techniques. However, as indicated in \cite{Allen}, BLAT is a popular algorithm being used in several researches to align probe sequences to a recent release of the human genome or transcriptome. The main advantage of the method is its low computational cost; BLAT does not keep the whole genome information in memory allowing high performance in an ordinary Linux computer. In addition, BLAT can be accessed for small searches through a web server at \, \texttt{\scriptsize https://genome.ucsc.edu/cgi-bin/hgBlat}. The sources and executables to install and run large mapping jobs in a computer are freely available for academic, personal and nonprofit purposes; see \; \texttt{\scriptsize http://www.kentinformatics.com} \; for details.

The results obtained via BLAT may contain some inconsistencies; a probe may be aligned to multiple locations and a location may be associated with two or more probes. We have no additional information to correct this mapping problem, which affects very few probes and locations; therefore, the involved expression values will be removed from the study without compromising the analysis.

In the data preprocessing routine, the intensity values are transformed using the standard $\log_2$-base scale and these measurements are adjusted within each chip and across the replicate arrays. Different sources can cause distortions in the data, for example, cross-hybridization, dust, chip defect, the amount of RNA in the samples, camera exposure time, scanner calibration, etc. The first three sources affect observations within a single chip and the remaining ones introduce noise between chips.

Among the different techniques to preprocess the data, we consider the Robust Multi-chip Average (RMA) \citep[see][]{IrizarryRMA}. It has three main steps: background correction, quantile normalization and summarization. In short, the first step fits a linear model ``signal plus error" to explain the intensity of probe $i$ in the array $j$; here, the main goal is to estimate the signal component. The quantile normalization step adjusts the probe intensities to ensure that measurements from different arrays are comparable. Finally, the summarization step calculates a single value representing the expressions of the probes in each probe set.

In our study, we apply only the first two RMA preprocessing steps. The third step is ignored because BLAT aligns the probe level data and not the probe set summarised measurement. In order to apply the RMA background correction and quantile normalization, we use the software \texttt{R} \citep{softwareR} and its package \texttt{affy} \citep{Gautier} integrated into the collaborative project Bioconductor (\texttt{\scriptsize http://www.bioconductor.org}) providing tools for computational biology \citep{Gentleman}. We also highlight the fact that both RMA and BLAT work only with PM probes; therefore, the MM probes are not considered in our final aligned data.

\newpage

\renewcommand{\thefigure}{B.\arabic{figure}} \setcounter{figure}{0}
\renewcommand{\theequation}{B.\arabic{equation}} \setcounter{equation}{0}
\renewcommand{\thetable}{B.\arabic{table}} \setcounter{table}{0}
{\flushleft \textbf{Appendix B: Joint and full conditional distributions} }
\vspace{10pt}

The joint density of $X$ and all the unknown quantities of the model is given by:
{\small
\begin{eqnarray}
\ds & &\pi(X,Z,W,V,q_0,Q,\psi,\beta) = \nonumber \\
    & = &\left[\prod_{i=1}^n\pi(X_i|Z_i,\psi)\pi(Z_i|Z_{i-1},W_i,q_0,Q)\pi(W_i|V_i)\pi(V_i|\beta)\right]
    \pi(q_0,Q,\psi,\beta)  \nonumber \\
    & = & \Bigg[ \prod_{i=1}^n \left[ \prod_{k=1}^{K+1} f_k(X_i|\psi)^{Z_{i,k}} (q_{0k}^{Z_{i,k}})^{1-W_i}(q_{k_{(i-1)}k}^{Z_{i,k}})^{W_i} \right] \nonumber \\
    &   & \times \; \left[ \mathds{1}(W_i = 1) \; \mathds{1}(V_i > 0) + \mathds{1}(W_i = 0) \; \mathds{1}(V_i \leq 0) \right] \nonumber \\
    &   & \times \; \phi(V_i-\beta'\vec{d_i}) \Bigg] \left[ \prod_{k=1}^{K+1} q_{0k}^{Z_{0,k}}q_{0k}^{r_{0k}-1} \right] \; \left[ \prod_{k_1=1}^{K+1} \prod_{k=1}^{K+1}q_{k_1 k}^{r_{k_1 k}-1} \right] \; \pi(\psi) \; \pi(\beta) \label{jdens},
	  \label{joidist}
\end{eqnarray}}
where \; $k_{(i-1)}=j$ \; if \; $Z_{i-1,j}=1$, \; $\pi(\beta)=|\Sigma_0|^{-1/2}\phi_2[\Sigma_{0}^{-1/2}(\beta-\mu_0)]$ \; and

{\small
\begin{eqnarray}
 \pi(\psi) & = & \ds \left[ \prod_{k=1}^{K}\pi_{IG}(\theta_k;t_{1k},t_{2k}) \; \mathds{1}(\theta_1<\ldots<\theta_K) \; \pi_{G}(\eta_k;e_{1k},e_{2k}) \right] \nonumber \\
           &   & \times \left[\pi_{NG}(\mu,\sigma^2;m,v,s_1,s_2) \; \mathds{1}(\mu > \max_{1\leq k\leq K}\{\theta_k\})\right]. \nonumber
\end{eqnarray}}

The full conditional distribution of $(q_0, Q, \psi, \beta)$ is:
\begin{eqnarray}
  (q_0|\cdot) &\sim& \mbox{Dir} \left[ r_{0} + \sum_{i=1}^n Z_{i} (1-W_i) \right]; \label{fc1}\\
  (q_k|\cdot) &\sim& \mbox{Dir} \left[ r_{k} + \sum_{i=2}^n (Z_{i-1,k} W_i) Z_{i} \right]; \label{fc2}\\
  (\beta|\cdot) &\sim& N_2(\mu_0^*,\Sigma_0^*),\label{fc3}
\end{eqnarray}
with $\Sigma_0^* = \left( \Sigma_0^{-1} + \sum_{i = 1}^n \vec{d}_i \vec{d_i}' \right)^{-1}$ and \; $\mu_0^* = \Sigma_0^* \left( \Sigma_0^{-1} \mu_0 + \sum_{i=1}^n V_i \vec{d}_i \right)$.

\begin{equation}
  (\sigma^2|\cdot) \sim IG(s_1^*,s_2^*); \;\;\;\; (\mu|\cdot) \sim N(m^*,v^*); \;\;\;\; (\theta_k|\cdot) \sim IG(t_{1k}^*,t_{2k}^*), \; k=1,\ldots,K; \label{fc4}
\end{equation}
where $s_1^* = s_1 + \frac{1}{2} \left( \sum_{i=1}^n Z_{i,K+1} \right)$ \; and \\
$s_2^* = s_2 + \frac{1}{2} \left[ \frac{m^2}{v} +
\sum_{i=1}^n Z_{i,K+1} X_i^2 -\frac{v}{1+v \sum_{i=1}^n Z_{i,K+1}} \left( \frac{m}{v} + \sum_{i=1}^n Z_{i,K+1} X_i \right)^2 \right]$; \\
$v^* = \frac{v \sigma^2}{1 + v \sum_{i=1}^n Z_{i,K+1}}$ \;  and \; $m^* = v^* \left( \frac{m + v \sum_{i=1}^n Z_{i,K+1} X_i}{v \sigma^2} \right)$;\\
$t_{1k}^* = t_{1k} + \eta_k \sum_{i=1}^n Z_{i,k}$ \; and \; $t_{2k}^* = t_{2k} + \eta_k \sum_{i=1}^n Z_{i,k} X_i$.\\

The MH algorithm to sample $\eta_k$ takes into account:
\begin{itemize}
\item Proposal distribution: \\
      \begin{equation}\label{MH1}
          \eta_{k}^{*}\sim N(\eta_{k},\tau^2), \;\; \mbox{where $\eta_{k}$ is the current value of the chain.}
      \end{equation}
\item Acceptance probability: \\
      \begin{equation}\label{MH2}
         \min\left\{1\;,\;\frac{\prod_{i;\;Z_{i,k}=1}f_k(X_i|\eta_{k}^*,\theta_k)\pi(\eta_{k}^*)}{\prod_{i;\;Z_{i,k}=1}f_k(X_i|\eta_{k},\theta_k)\pi(\eta_{k})}\right\};
      \end{equation}
      where $\theta_{k}$ is the current value of parameter $\theta_k$.
\end{itemize}

The sampling step of $(Z,W)$ is as follows:
\begin{eqnarray}
 Z_1  & \sim & \mbox{Mult}(1,q_{0}^*); \nonumber \\
 (W_i|Z_{i-1,j}=1,\cdot) & \sim & \mbox{Ber}(1,p_{(i,j)}^*),\;i=2,\ldots,n; \nonumber \\
 (Z_i|W_i=l,Z_{i-1,j}=1,\cdot) & \sim & \mbox{Mult}(1,q_{(i,j,l)}^*),\;i=2,\ldots,n \nonumber
\end{eqnarray}
where \vspace{-3pt}
\begin{eqnarray} \label{fsampling}
 q_{0k}^* &=& c_{1,k}q_{0k}/a_1; \\
 p_{(i,j)}^* &=& b_{i,j}\Phi_{i}^+/c_{i,j}; \nonumber \\
 q_{(i,j,0)k}^* &=& c_{i+1,k}f_k(X_i|\psi)q_{0k}/a_i;\;\;\;q_{(i,j,1)k}^*  = c_{i+1,k}f_k(X_i|\psi)q_{j,k}/b_{i,j}. \nonumber
\end{eqnarray}
Consider $c_{n+1,j}=1$, \, $\forall j$, \, and:
\begin{eqnarray} \label{filtering}
  a_n & = & \sum_{k=1}^{K+1}f_k(X_n|\psi)q_{0k}; \;\;\; b_{n,j} = \sum_{k=1}^{K+1}f_k(X_n|\psi)q_{j,k}; \;\;\; c_{n,j}=b_{n,j}\Phi_{n}^++a_n\Phi_{n}^- ; \\
  a_{i} & = & \sum_{k=1}^{K+1}c_{i+1,k}f_k(X_i|\psi)q_{0k}; \;\; b_{i,j} = \sum_{k=1}^{K+1}c_{i+1,k}f_k(X_i|\psi)q_{j,k}; \;\; c_{i,j} = b_{i,j}\Phi_{i}^++a_i\Phi_{i}^-, \nonumber \\
  & & \mbox{for }i=n-1,\ldots,2; \nonumber \\
  a_{1} & = & \sum_{k=1}^{K+1}c_{1,k}q_{0k}. \nonumber
\end{eqnarray}
In the filtering procedure, the calculations are performed recursively, starting from $n$ and moving backwards. On each step, we compute the scalar $a_i$ and the $(K+1)$-vectors $b_i$ and $c_i$ as indicated in (\ref{filtering}). Next, we compute the probabilities in (\ref{fsampling}) to update the auxiliary variables $Z_i$ and $W_i$.

\newpage

\vspace{5pt}
\renewcommand{\thefigure}{C.\arabic{figure}} \setcounter{figure}{0}
\renewcommand{\theequation}{C.\arabic{equation}} \setcounter{equation}{0}
\renewcommand{\thetable}{C.\arabic{table}} \setcounter{table}{0}
{\flushleft \textbf{Appendix C: Additional results of the real data application} }
\vspace{10pt}

\begin{figure}[!h]
$$
  \begin{array}{cc}
  \hspace{-0.2cm} \mbox{\scriptsize Chromosome 2} & \hspace{-0.5cm} \mbox{\scriptsize Chromosome 3} \\
  \hspace{-0.2cm} \includegraphics[scale=0.18]{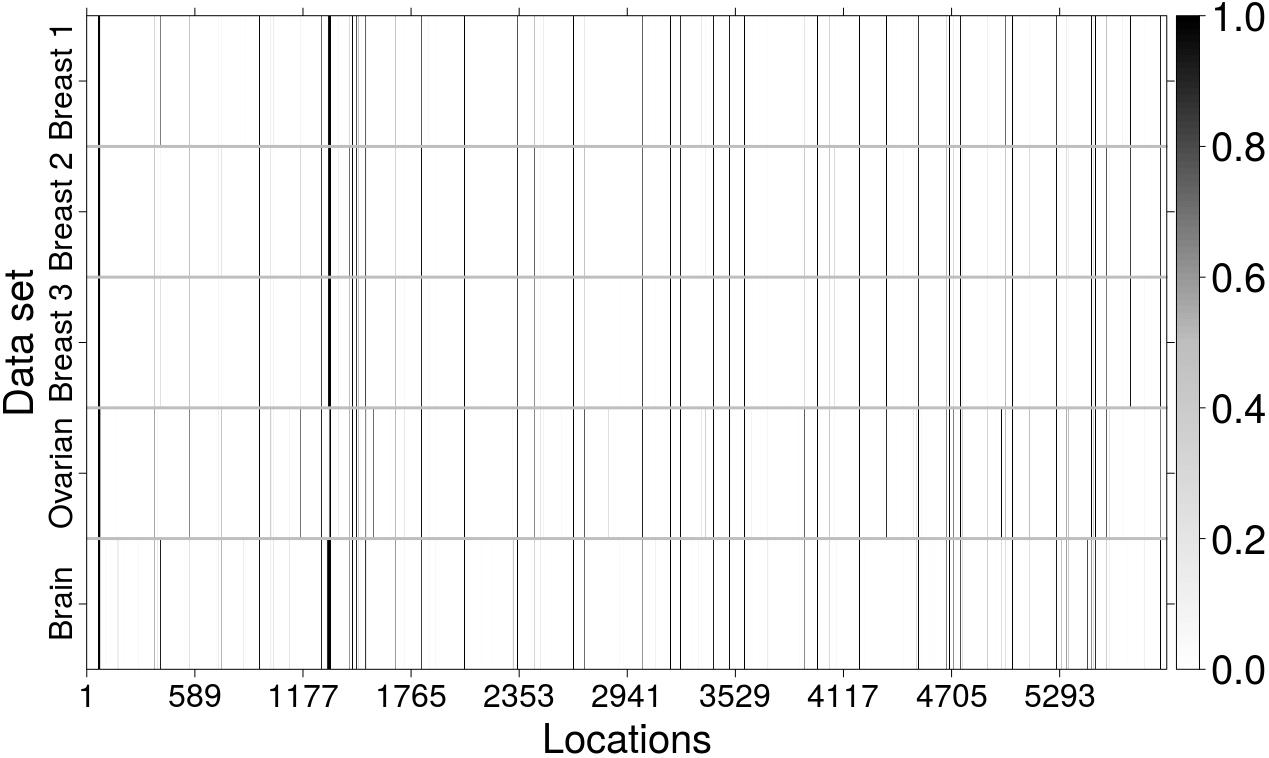} & \hspace{-0.2cm} \includegraphics[scale=0.18]{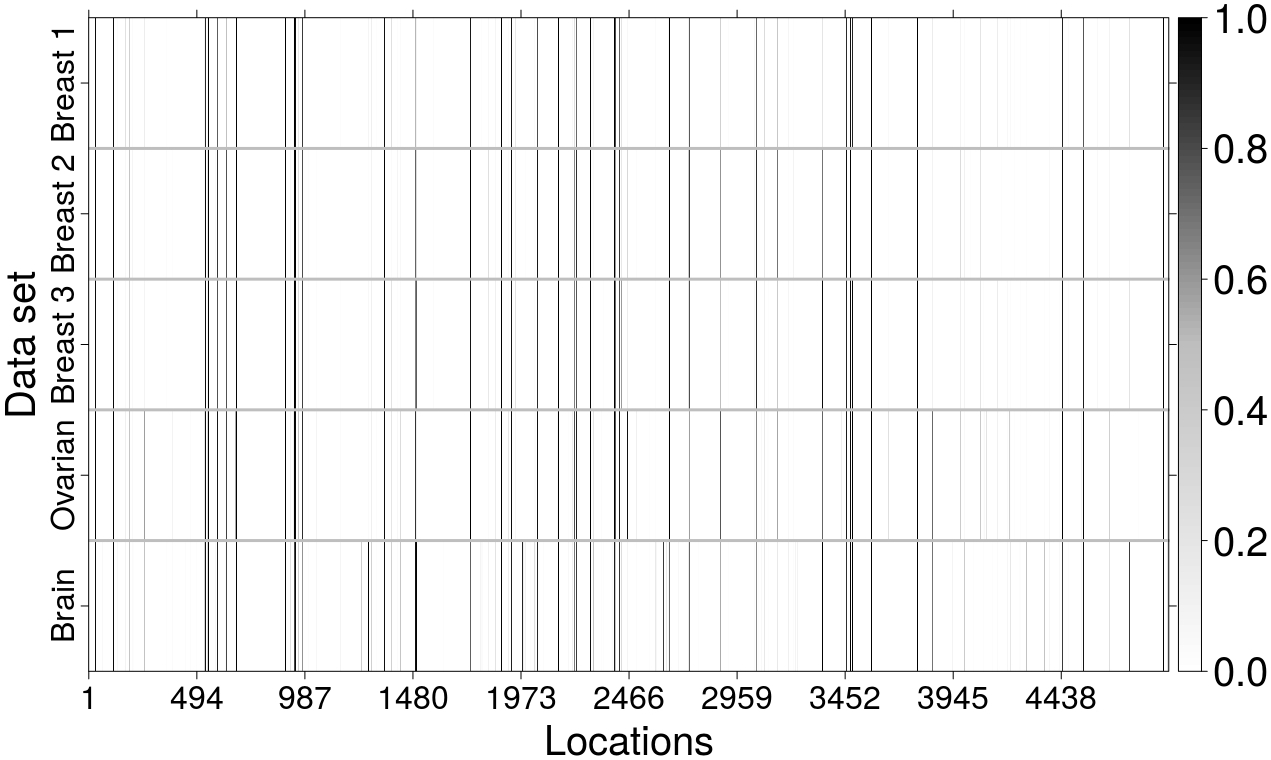}\\
  \hspace{-0.2cm} \mbox{\scriptsize Chromosome 4} & \hspace{-0.5cm} \mbox{\scriptsize Chromosome 5} \\
  \hspace{-0.2cm} \includegraphics[scale=0.18]{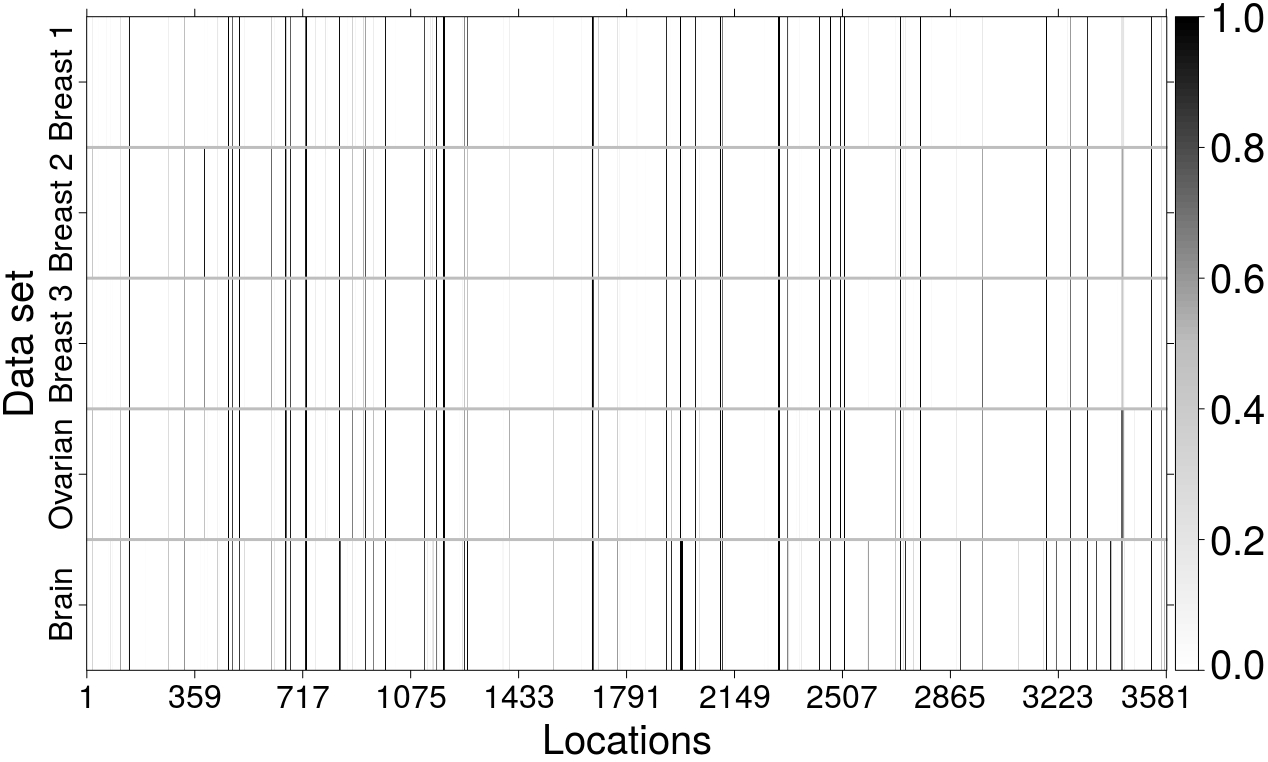} & \hspace{-0.2cm} \includegraphics[scale=0.18]{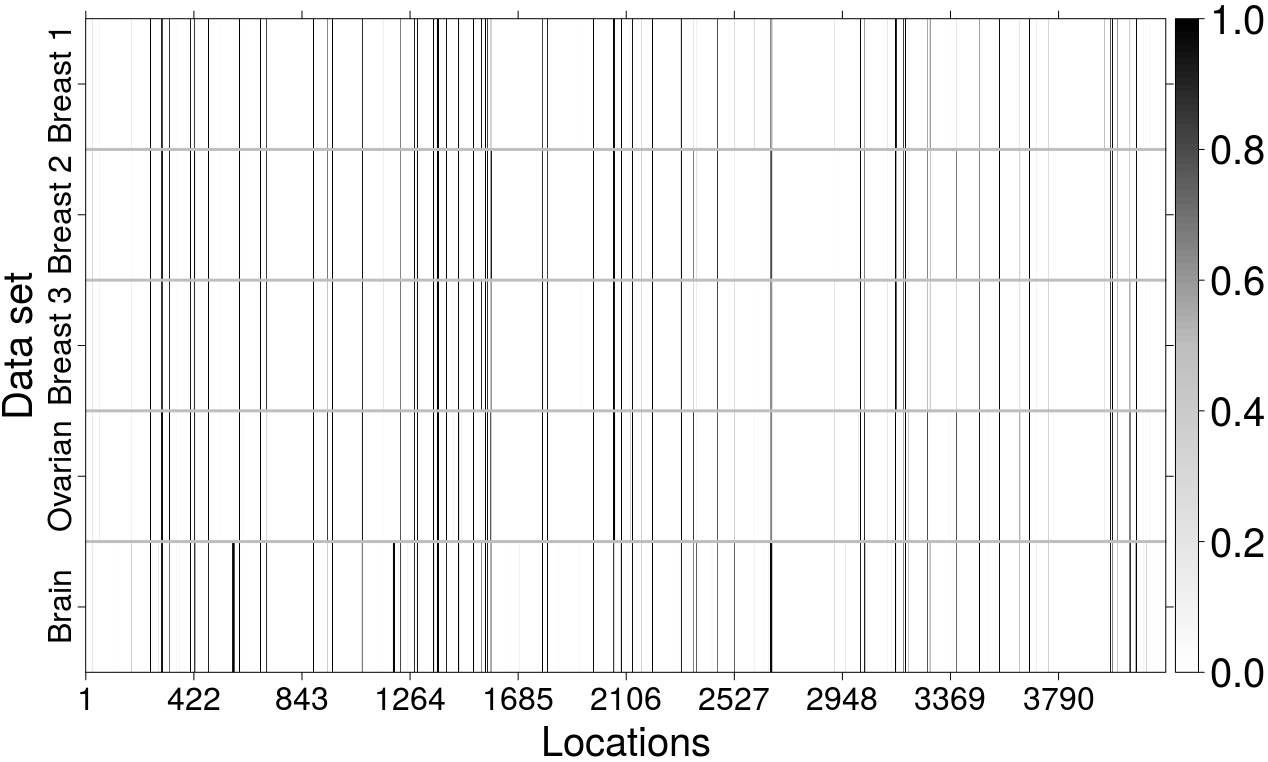}\\
  \hspace{-0.2cm} \mbox{\scriptsize Chromosome 6} & \hspace{-0.5cm} \mbox{\scriptsize Chromosome 7} \\
  \hspace{-0.2cm} \includegraphics[scale=0.18]{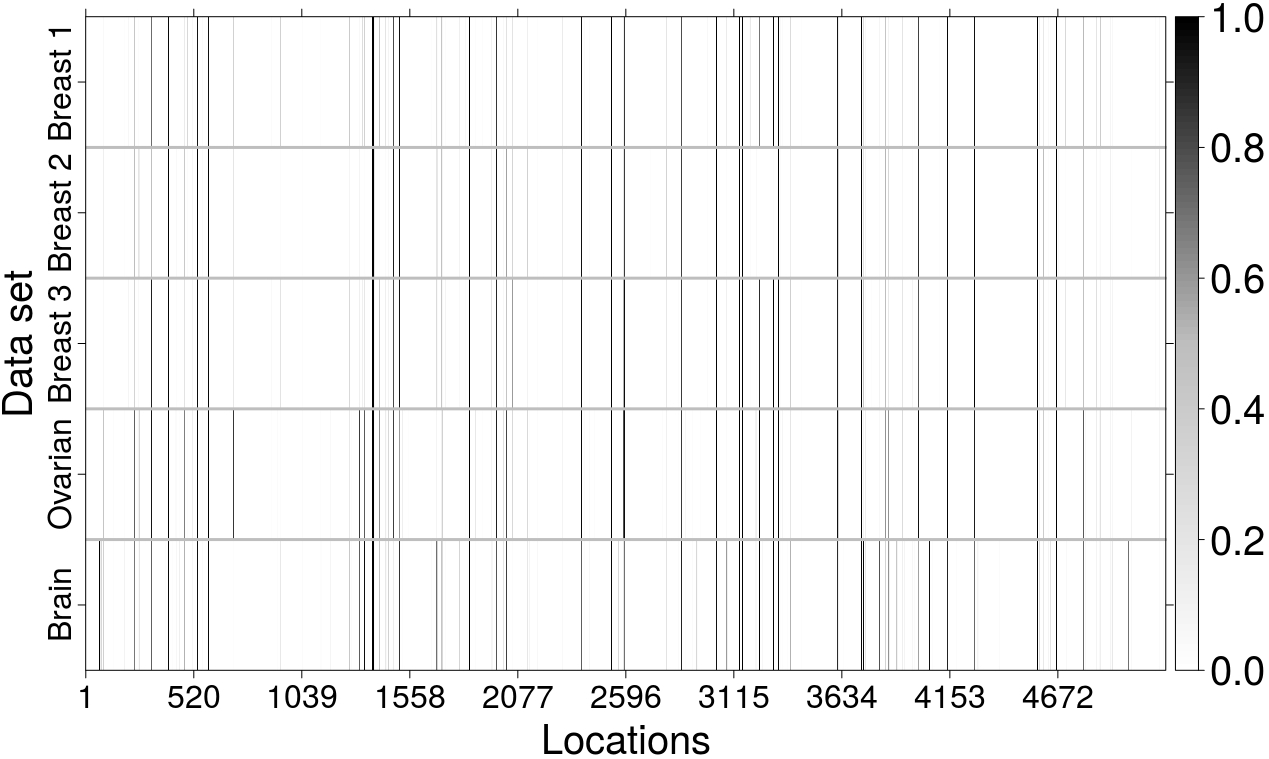} & \hspace{-0.2cm} \includegraphics[scale=0.18]{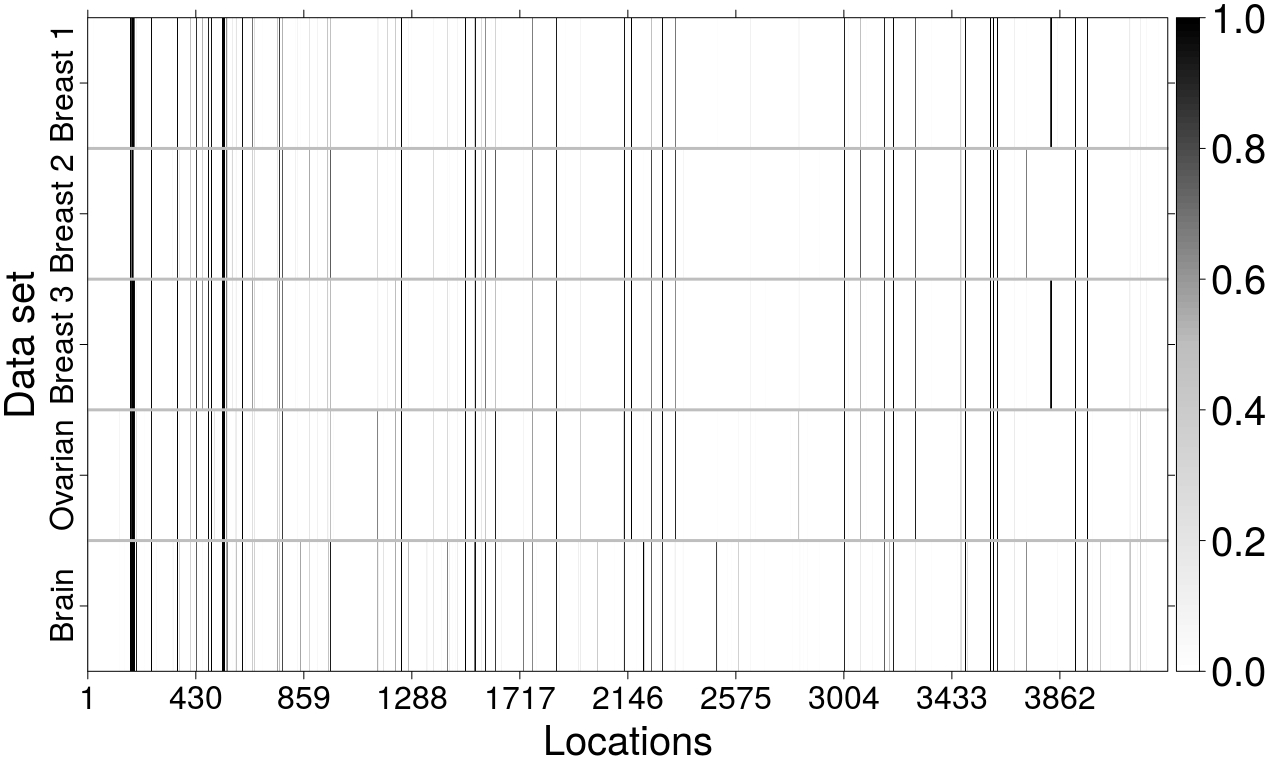}\\
  \hspace{-0.2cm} \mbox{\scriptsize Chromosome 8} & \hspace{-0.5cm} \mbox{\scriptsize Chromosome 9} \\
  \hspace{-0.2cm} \includegraphics[scale=0.18]{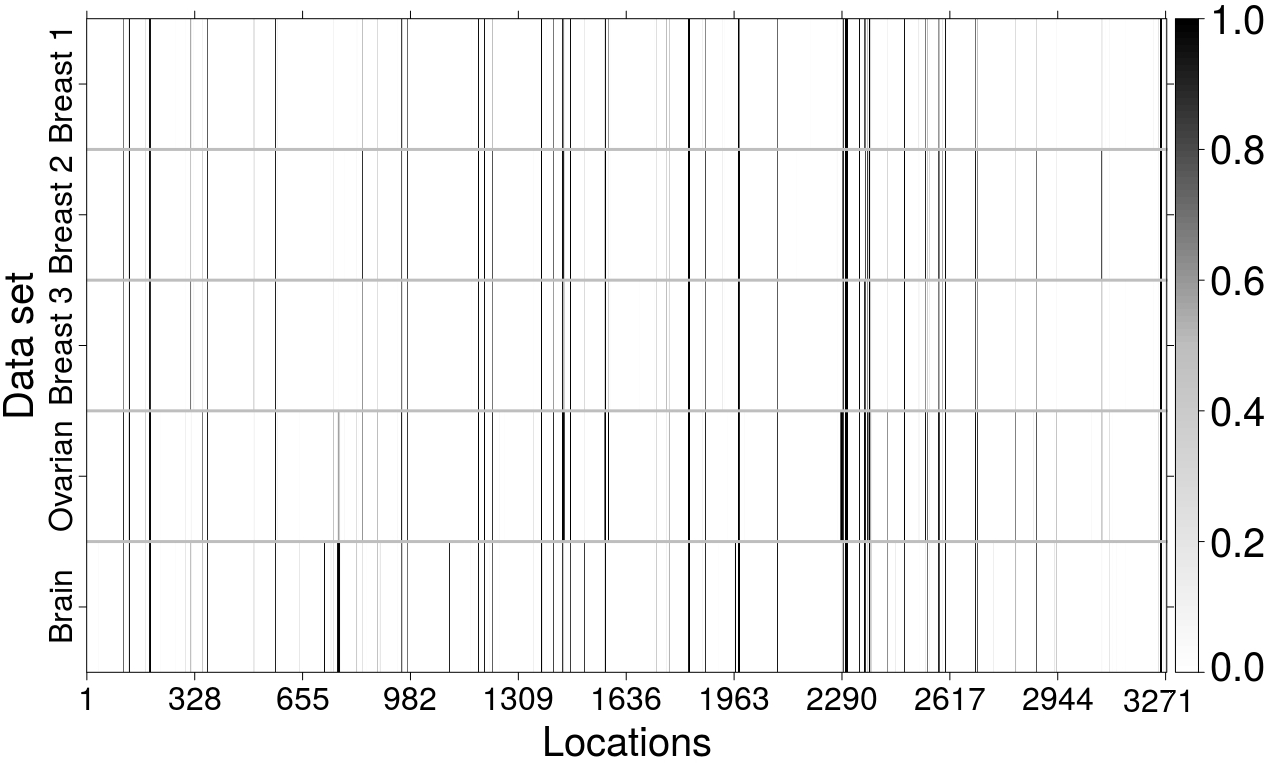} & \hspace{-0.2cm} \includegraphics[scale=0.18]{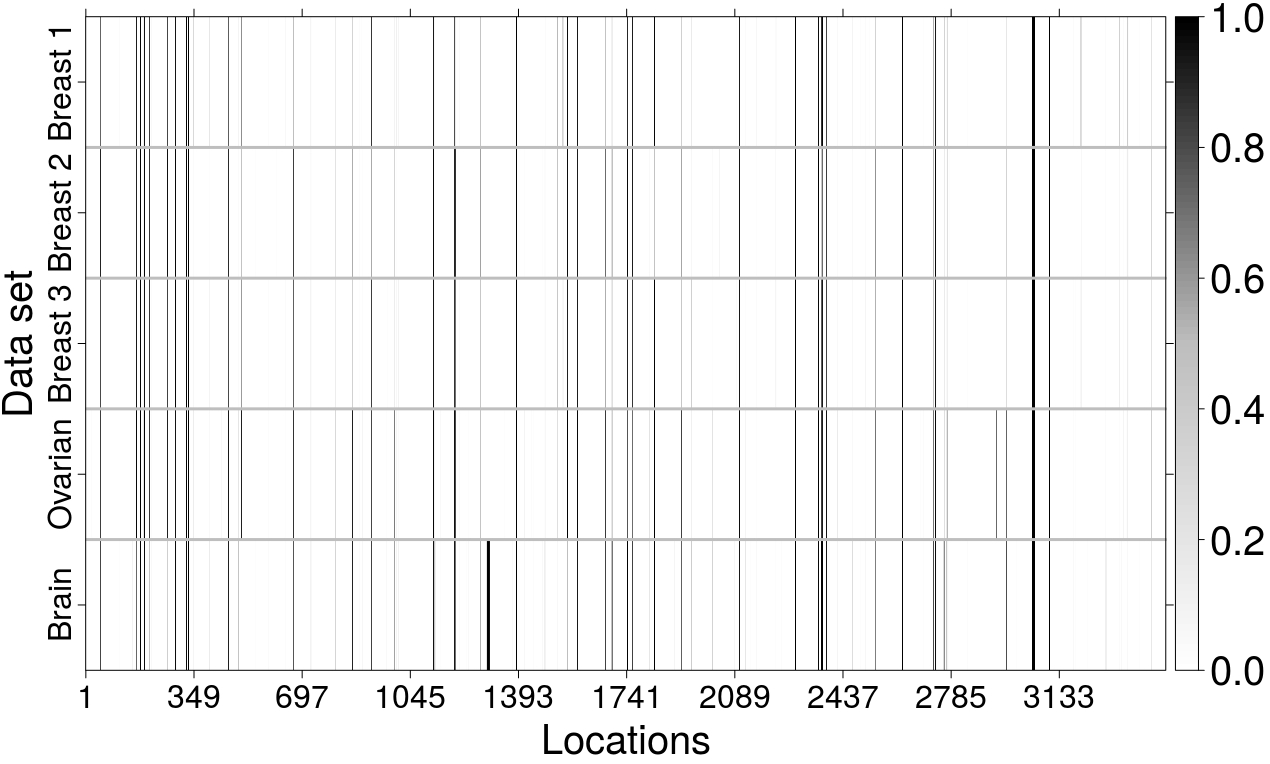}\\
  \end{array}
$$
\vspace{-15pt} \caption{\scriptsize Heat map image (chromosomes 2-9) indicating for each location the posterior probability of belonging to the Gaussian component.}
\label{figB1}
\end{figure}

\newpage

\begin{figure}[!h]
$$
  \begin{array}{cc}
  \hspace{-0.2cm} \mbox{\scriptsize Chromosome 10} & \hspace{-0.5cm} \mbox{\scriptsize Chromosome 11} \\
  \hspace{-0.2cm} \includegraphics[scale=0.18]{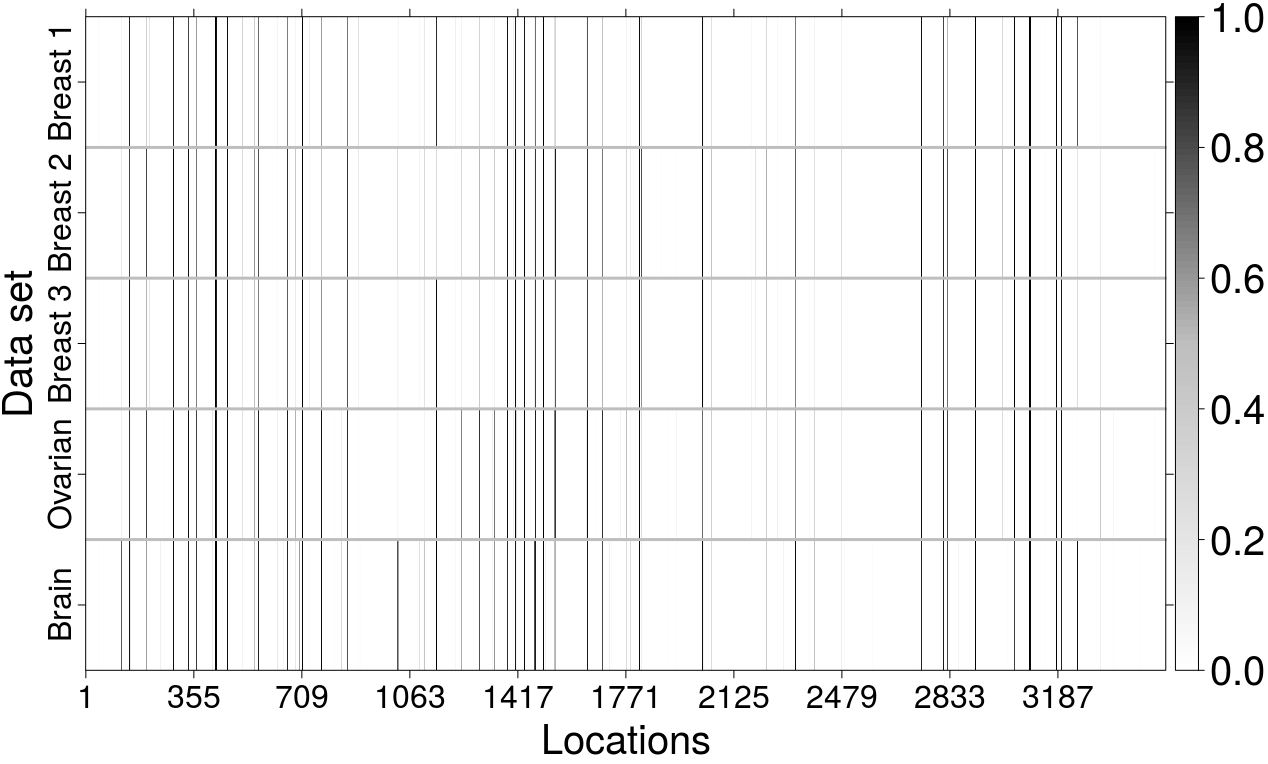} & \hspace{-0.2cm} \includegraphics[scale=0.18]{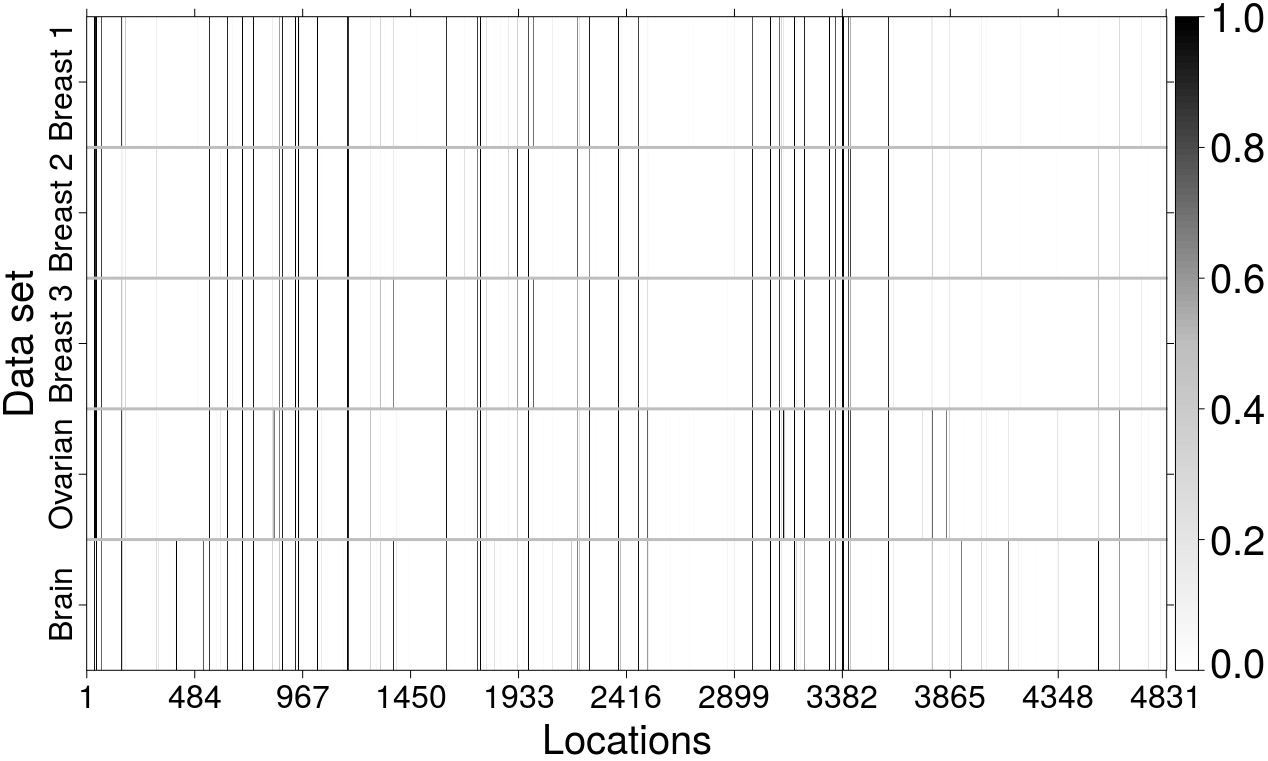}\\
  \hspace{-0.2cm} \mbox{\scriptsize Chromosome 12} & \hspace{-0.5cm} \mbox{\scriptsize Chromosome 13} \\
  \hspace{-0.2cm} \includegraphics[scale=0.18]{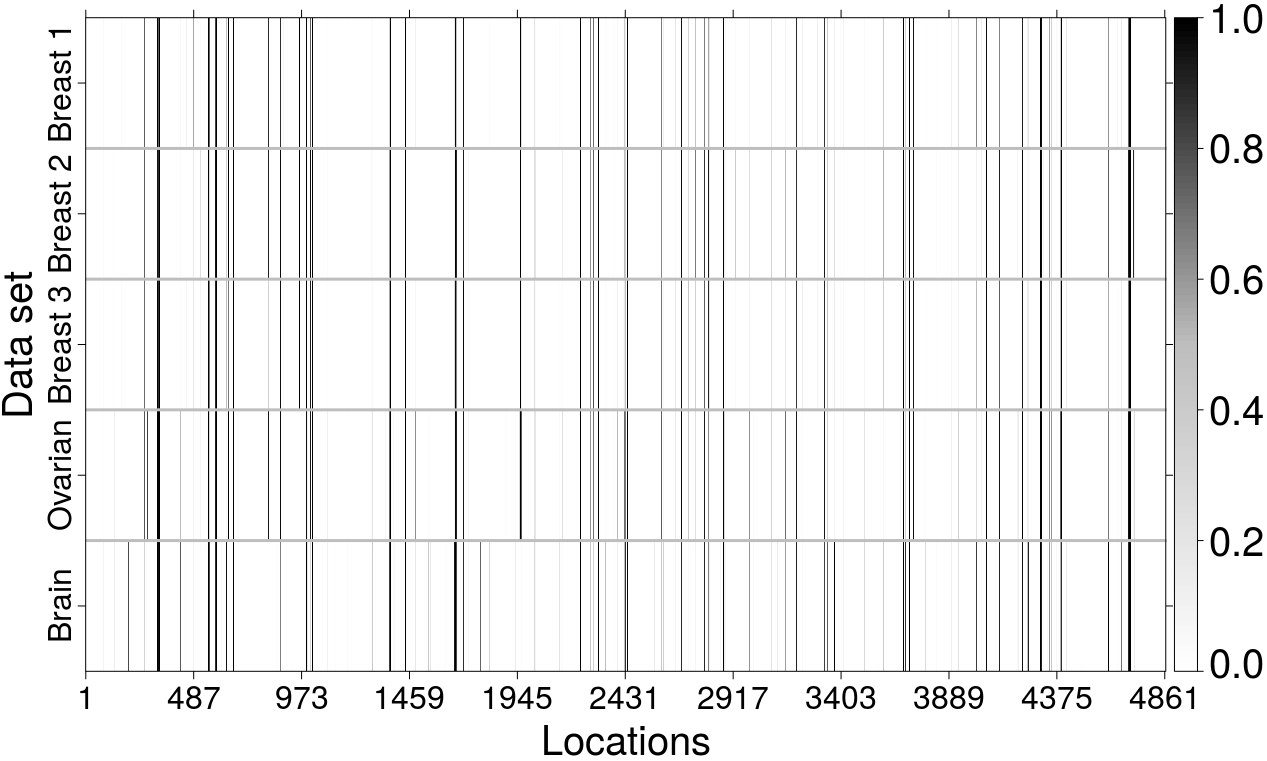} & \hspace{-0.2cm} \includegraphics[scale=0.18]{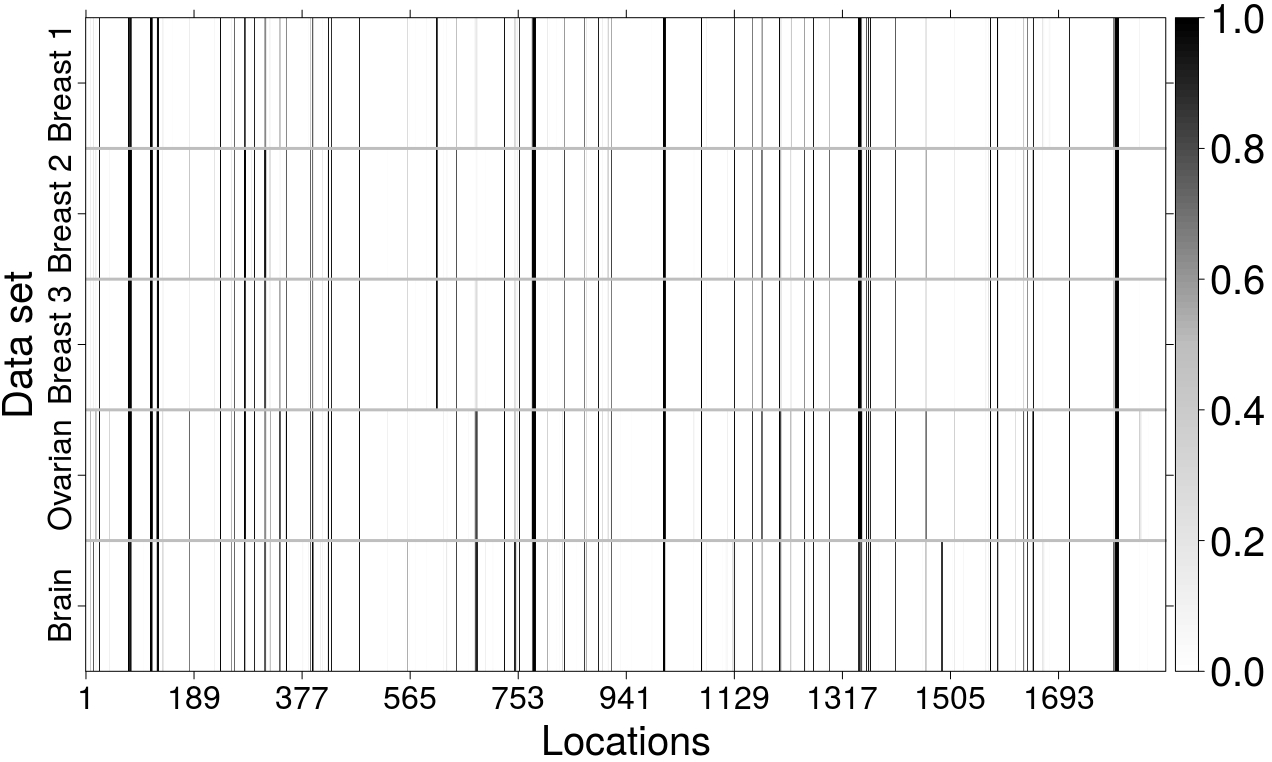}\\
  \hspace{-0.2cm} \mbox{\scriptsize Chromosome 14} & \hspace{-0.5cm} \mbox{\scriptsize Chromosome 15} \\
  \hspace{-0.2cm} \includegraphics[scale=0.18]{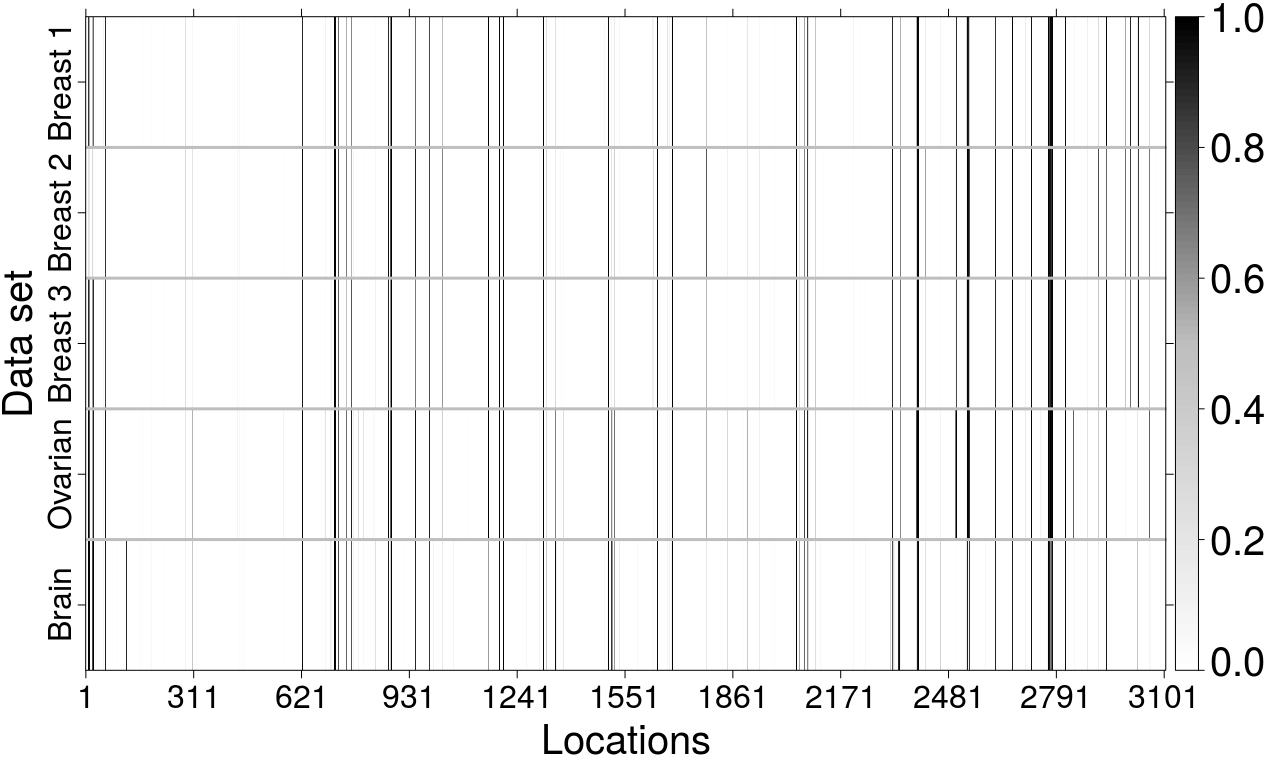} & \hspace{-0.2cm} \includegraphics[scale=0.18]{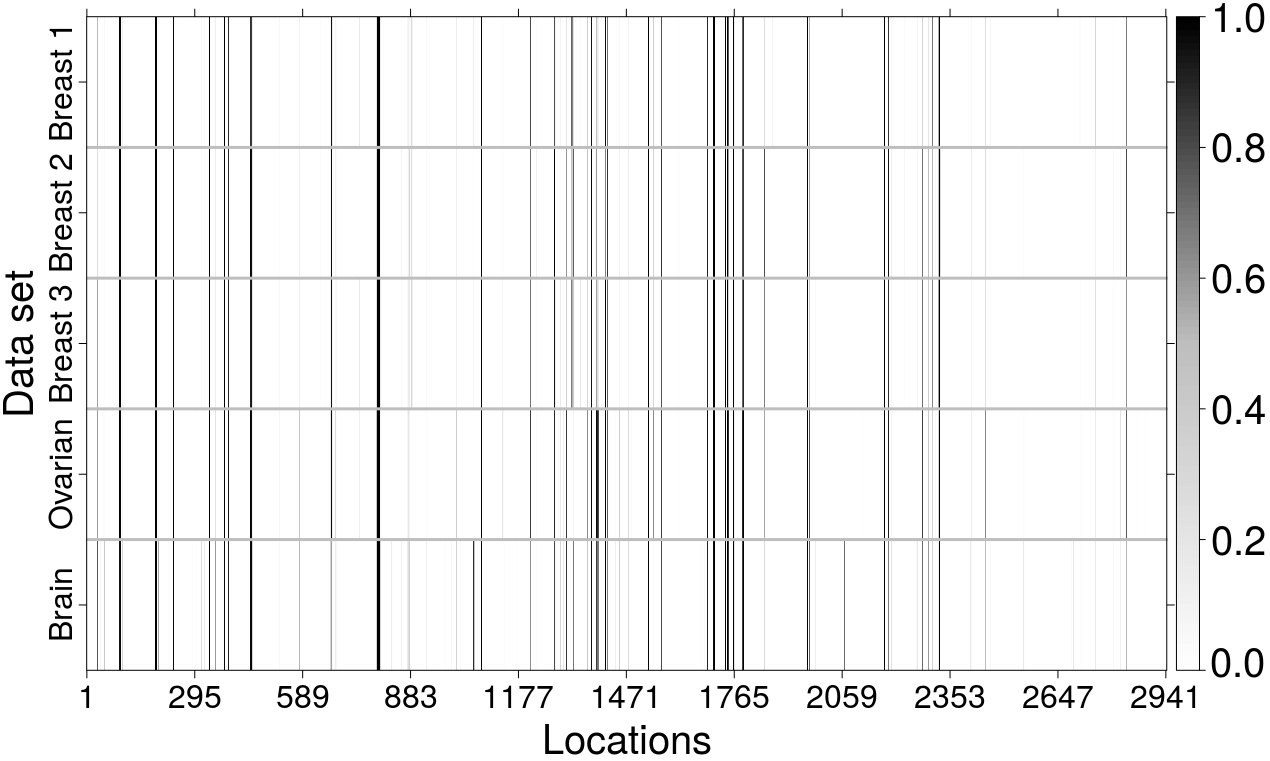}\\
  \hspace{-0.2cm} \mbox{\scriptsize Chromosome 16} & \hspace{-0.5cm} \mbox{\scriptsize Chromosome 17} \\
  \hspace{-0.2cm} \includegraphics[scale=0.18]{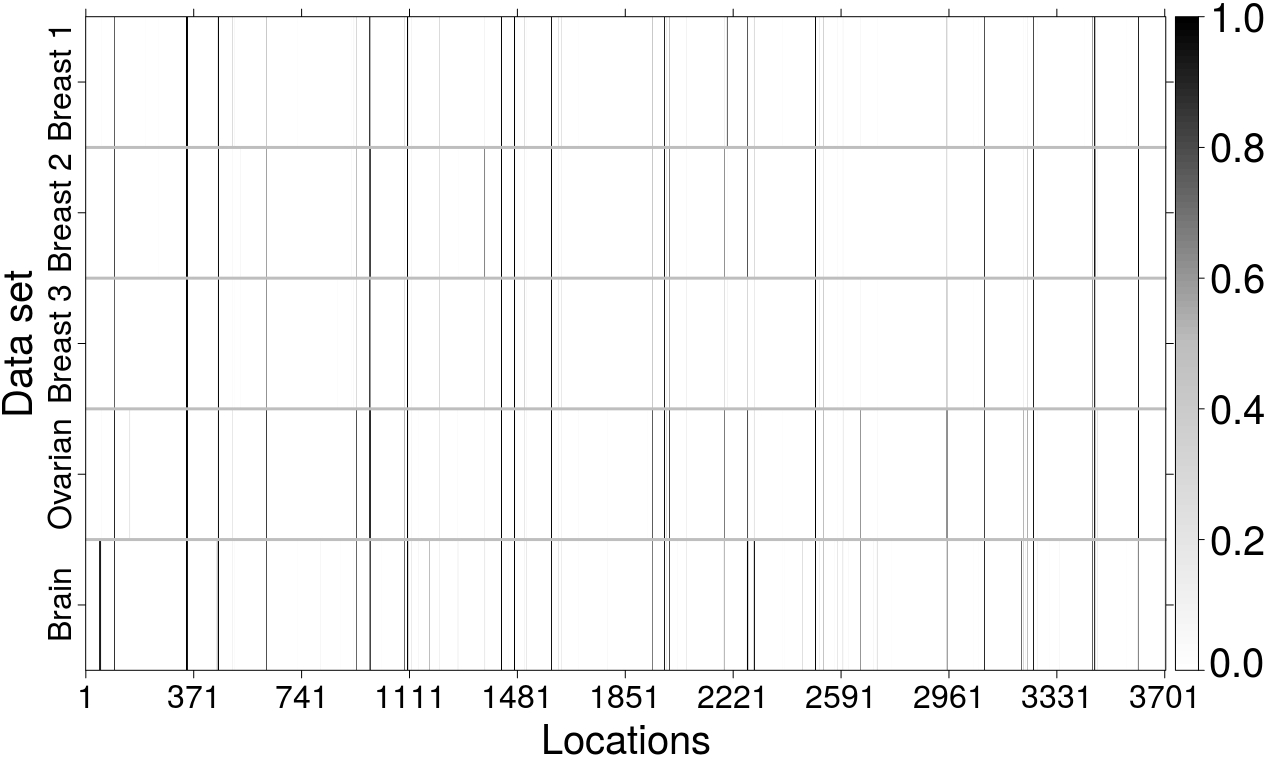} & \hspace{-0.2cm} \includegraphics[scale=0.18]{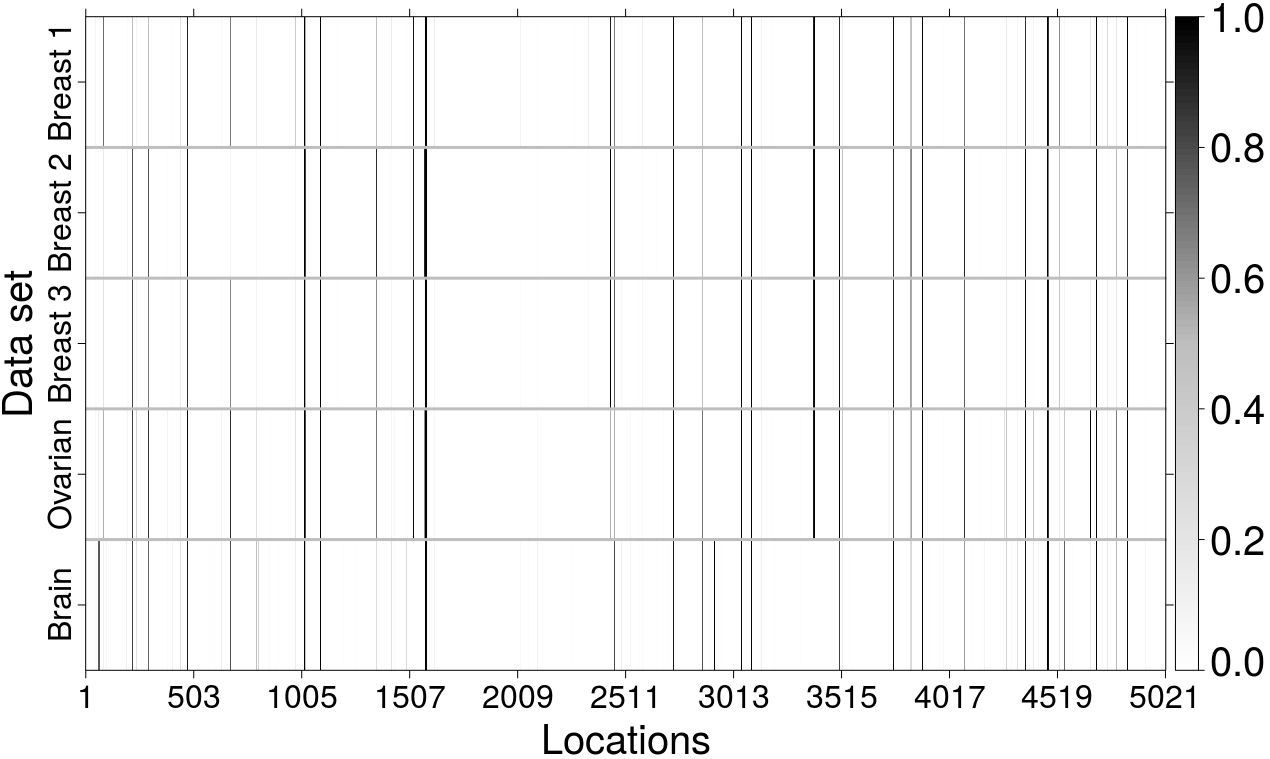}\\
	\end{array}
$$
\vspace{-15pt} \caption{\scriptsize Heat map image (chromosomes 10-17) indicating for each location the posterior probability of belonging to the Gaussian component.}
\label{figB2}
\end{figure}

\newpage

\begin{figure}[!h]
$$
  \begin{array}{cc}
  \hspace{-0.2cm} \mbox{\scriptsize Chromosome 18} & \hspace{-0.5cm} \mbox{\scriptsize Chromosome 19} \\
  \hspace{-0.2cm} \includegraphics[scale=0.18]{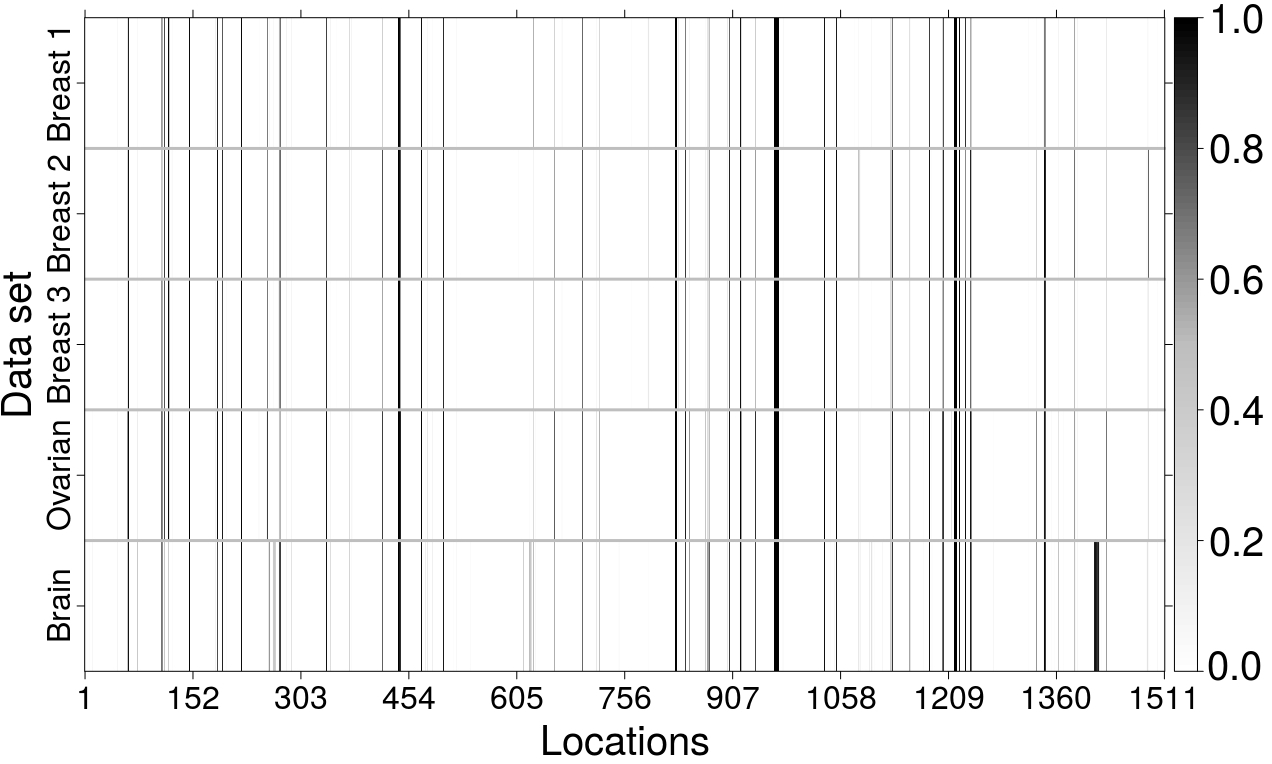} & \hspace{-0.2cm} \includegraphics[scale=0.18]{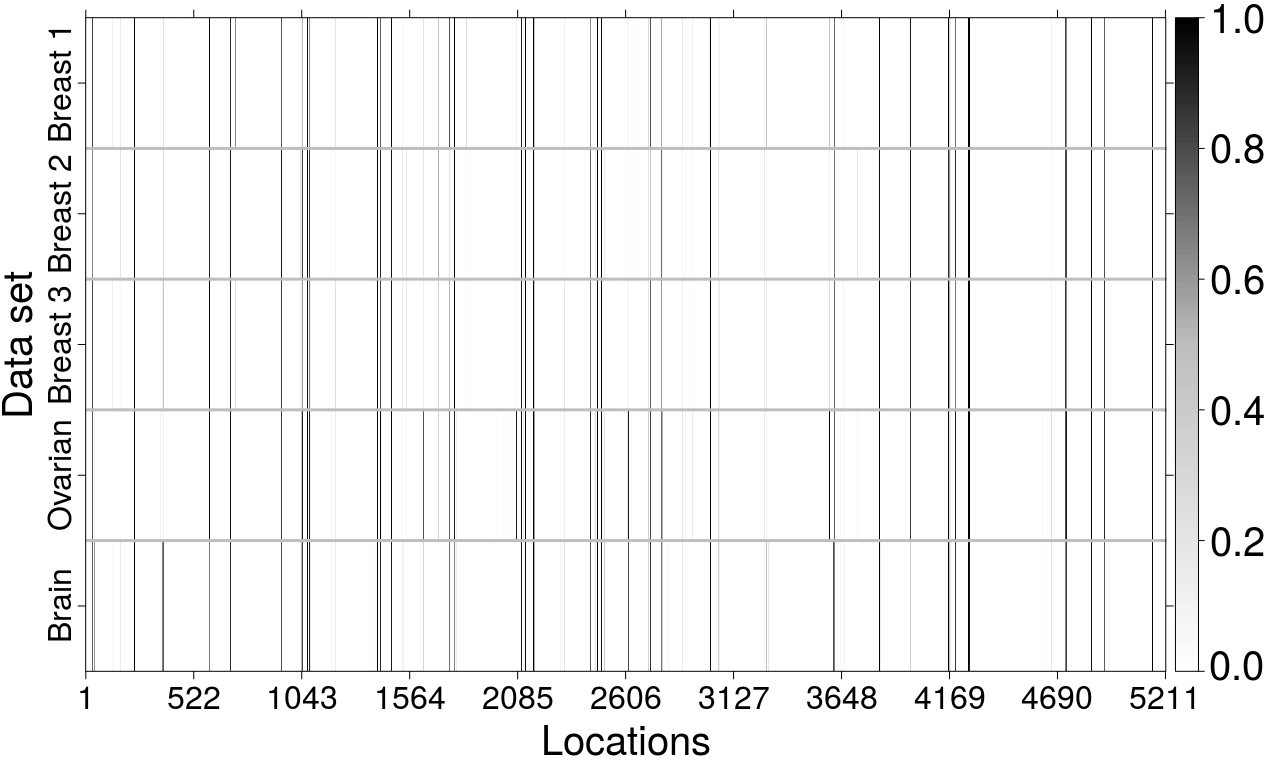}\\
  \hspace{-0.2cm} \mbox{\scriptsize Chromosome 20} & \hspace{-0.5cm} \mbox{\scriptsize Chromosome 21} \\
  \hspace{-0.2cm} \includegraphics[scale=0.18]{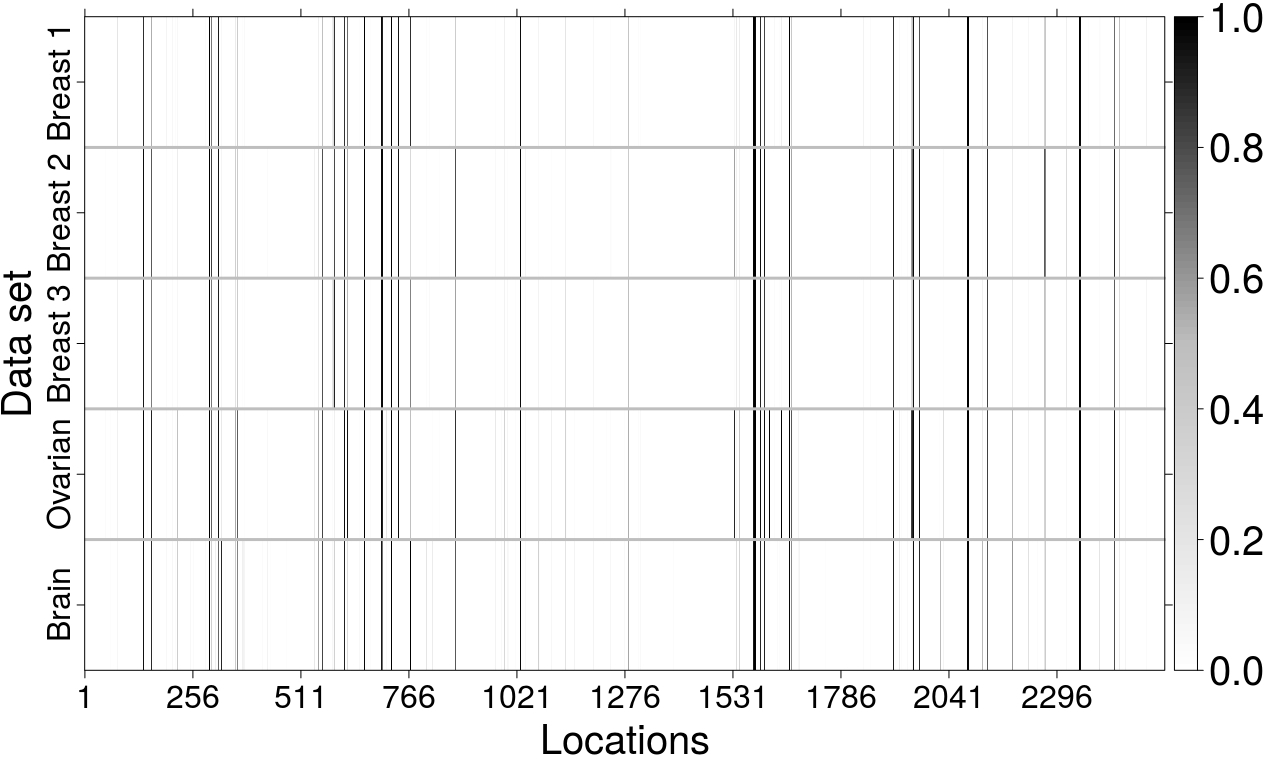} & \hspace{-0.2cm} \includegraphics[scale=0.18]{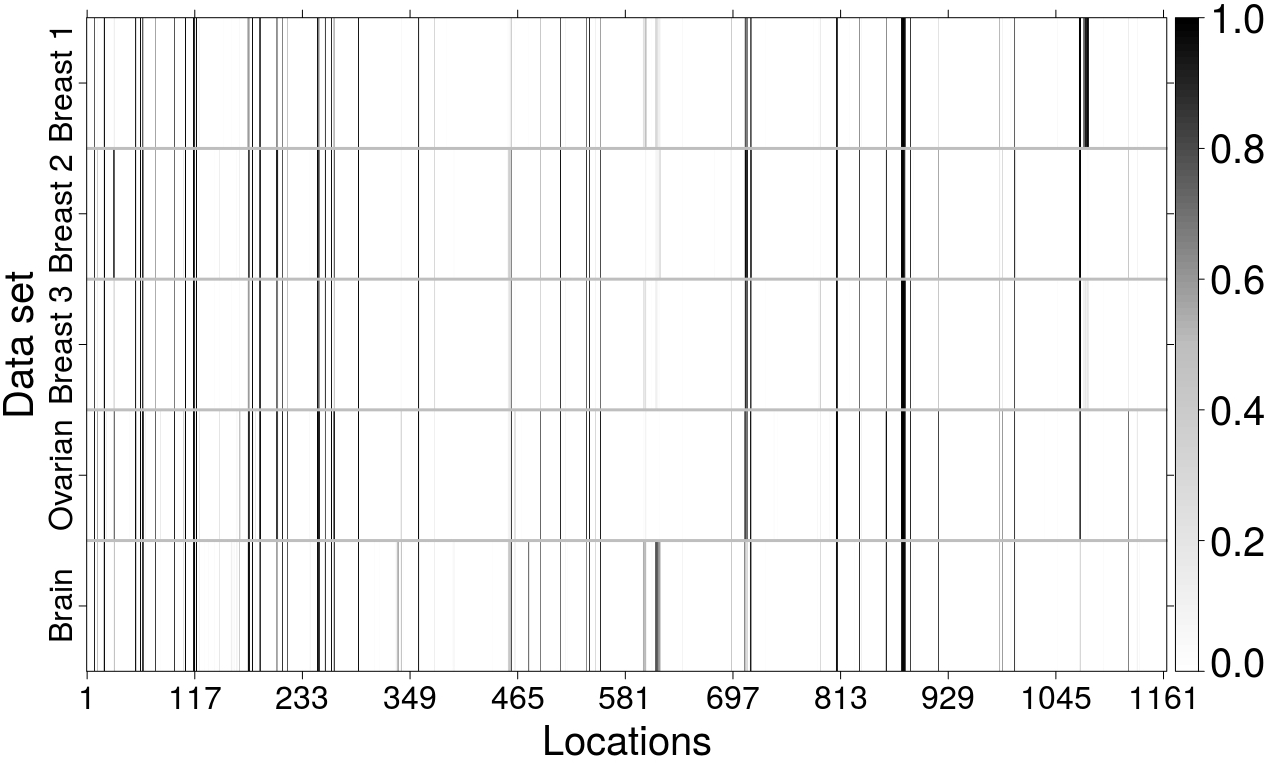}\\
  \hspace{-0.2cm} \mbox{\scriptsize Chromosome 22} & \hspace{-0.5cm} \mbox{\scriptsize Chromosome X} \\
  \hspace{-0.2cm} \includegraphics[scale=0.18]{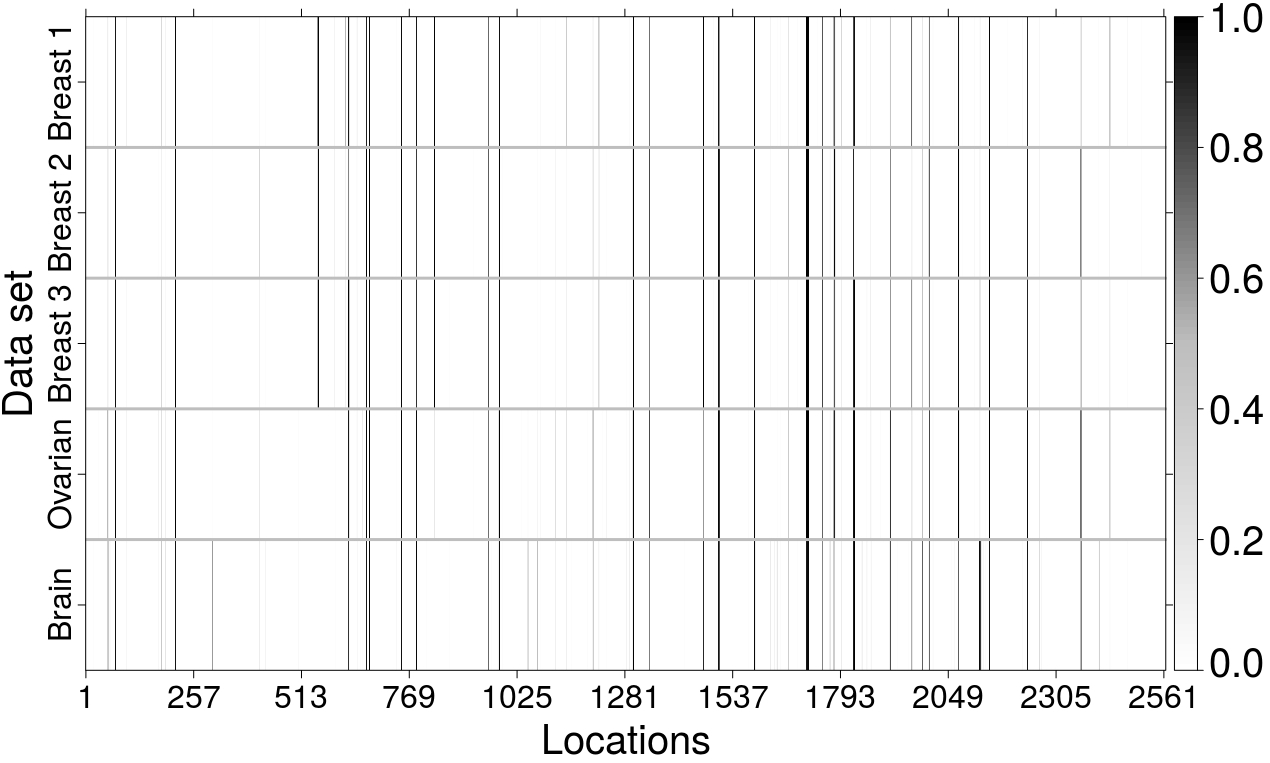} & \hspace{-0.2cm} \includegraphics[scale=0.18]{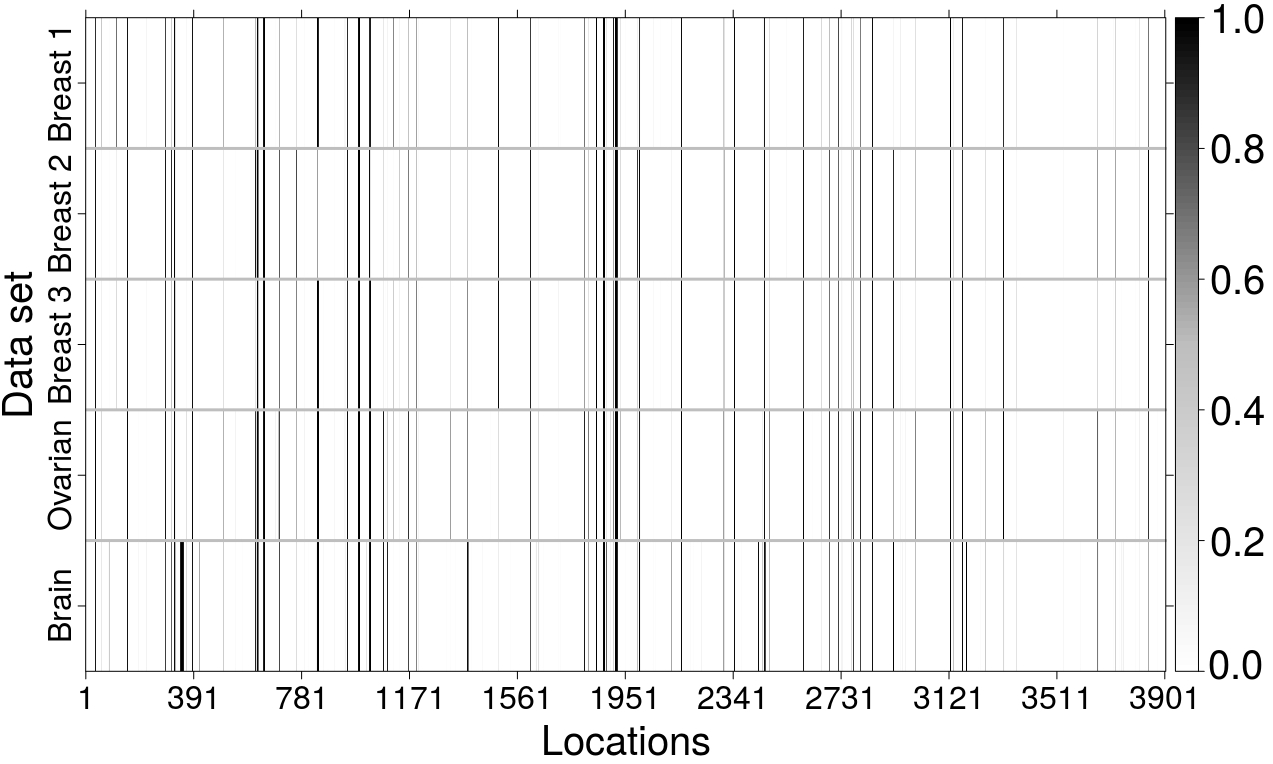}\\
  \hspace{-0.2cm} \mbox{\scriptsize Chromosome Y} & \\
  \hspace{-0.2cm} \includegraphics[scale=0.18]{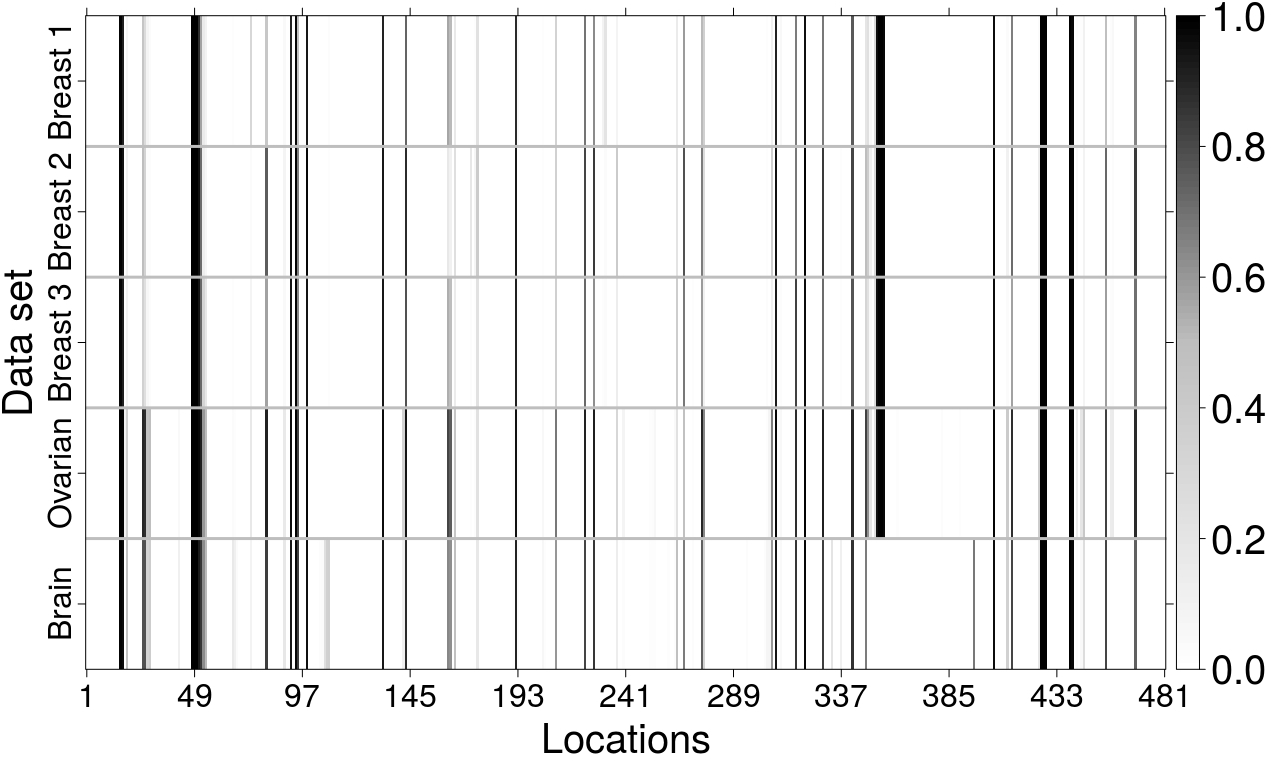} & \\
  \end{array}
$$
\vspace{-15pt} \caption{\scriptsize Heat map image (chromosomes 18-22, $X$ and $Y$) indicating for each location the posterior probability of belonging to the Gaussian component.}
\label{figB3}
\end{figure}

\vspace{5pt}
\renewcommand{\thefigure}{D.\arabic{figure}} \setcounter{figure}{0}
\renewcommand{\theequation}{D.\arabic{equation}} \setcounter{equation}{0}
\renewcommand{\thetable}{D.\arabic{table}} \setcounter{table}{0}
{\flushleft \textbf{Appendix D: Sensitivity analysis} }
\vspace{10pt}

\begin{figure}[!h]
\centering
   \includegraphics[scale=0.40]{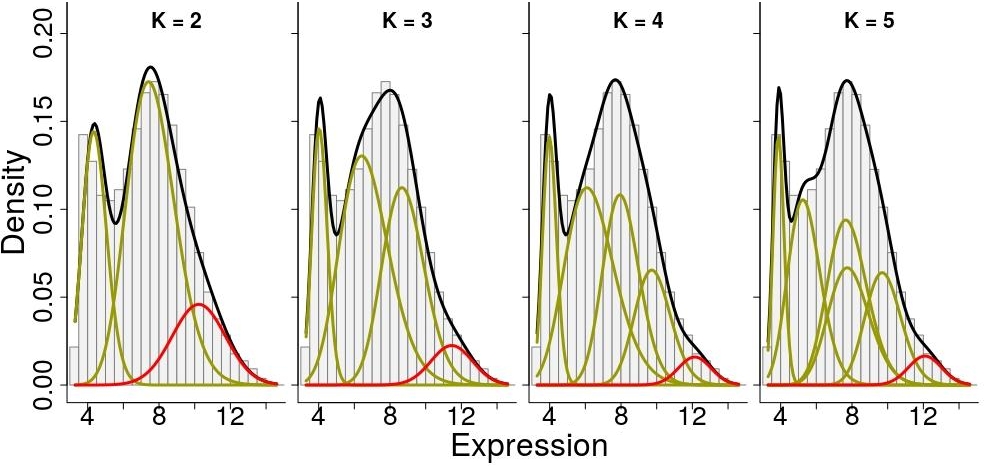}
\vspace{-10pt} \caption{\scriptsize Histogram of all expressions (``Breast 1'' data) overlaid by the estimated mixture density (black curve) and its components (gammas in yellow, normal in red), for all the four values of $K$.}
\label{SAK1}
\end{figure}

In order to investigate the sensitivity of the results to the choice of $K$ (number of gamma components), we fit the model under four different specifications: $K=2$, $3$, $4$ and $5$. The Dirichlet priors for $q_0$ and $Q$ are specified in a way that the Gaussian component's weight and the degree of information in the prior (sum of the hyperparameters of the Dirichlet) are the same across all configurations. Results are presented in Figure \ref{SAK1} and Table \ref{SAK2}. They indicate that $K=2$ does not provide good results as the Gaussian component's variance is too high to accommodate overexpressed probes. Results are quite similar (but significantly different) for $K=3$ and $K=4$, and they are practically the same for $K=4$ and $K=5$. In particular, two of the gamma components for $K=5$ are virtually the same -- the posterior mean of $(\theta,\eta)$ for each of these components are $(7.81,49.23)$ and $(7.90,46.89)$. Hence, we choose the model with $K=4$, for which results are reported in Section \ref{secresult}.

\begin{table}[!h]
\caption[caption]{\label{SAK2} \scriptsize Posterior estimates (for each $K$, ``Breast 1'' data) of the weight, $\mu$ and $\sigma^2$ related to the Gaussian component, and \\\hspace{\textwidth}
the mean (Expec.)  and variance (Var.) of the (normalised) mixture of gammas; standard errors in parentheses.}
{\scriptsize\begin{adjustbox}{center}
\fbox{%
\begin{tabular}{cccccc}
\noalign{\smallskip}
  $K$  & weight & $\mu$ & $\sigma^2$ & Expec. & Var. \\
 \hline
 \noalign{\smallskip}
 $2$ & 0.1715 (0.0033)   & 10.2320 (0.0281)   & 2.2208 (0.0362) & 6.6956 (0.0103) & 2.3649 (0.0177) \\
 \noalign{\smallskip}
 $3$ & 0.0654 (0.0017)   & 11.4605 (0.0341)   & 1.3443 (0.0388) & 7.0111 (0.0074) & 2.8614 (0.0209) \\
 \noalign{\smallskip}
 $4$ & 0.0365 (0.0011)   & 12.1317 (0.0330)   & 0.8314 (0.0338) & 7.1193 (0.0062) & 3.2746 (0.0226) \\
 \noalign{\smallskip}
 $5$ & 0.0383 (0.0012)   & 12.0889 (0.0354)   & 0.8599 (0.0354) & 7.1112 (0.0060) & 3.4991 (0.0204) \\
\end{tabular}}
\end{adjustbox}}
\end{table}

\begin{table}[!h]
\caption[caption]{\label{SAK3} \scriptsize Comparison of prior specifications (``Breast 1'' data). Posterior estimates of the weight, $\mu$ and $\sigma^2$ related to the Gaussian component, and the mean (Expec.) and variance (Var.) of the (normalised) mixture of gammas; standard errors in parentheses.}
{\scriptsize\begin{adjustbox}{center}
\fbox{%
\begin{tabular}{rccccc}
\noalign{\smallskip}
Prior & weight & $\mu$ & $\sigma^2$ & Expec. & Var. \\
\hline
\noalign{\smallskip}
Original & 0.0365 (0.0011) & 12.1317 (0.0330) & 0.8314 (0.0338) & 7.1193 (0.0059) & 3.2746 (0.0226) \\
\noalign{\smallskip}
Vague    & 0.0364 (0.0011) & 12.1338 (0.0333) & 0.8305 (0.0334) & 7.1197 (0.0059) & 3.2736 (0.0223) \\
\end{tabular}}
\end{adjustbox}}
\end{table}

\begin{figure}[!h]
\centering
   \includegraphics[scale=0.32]{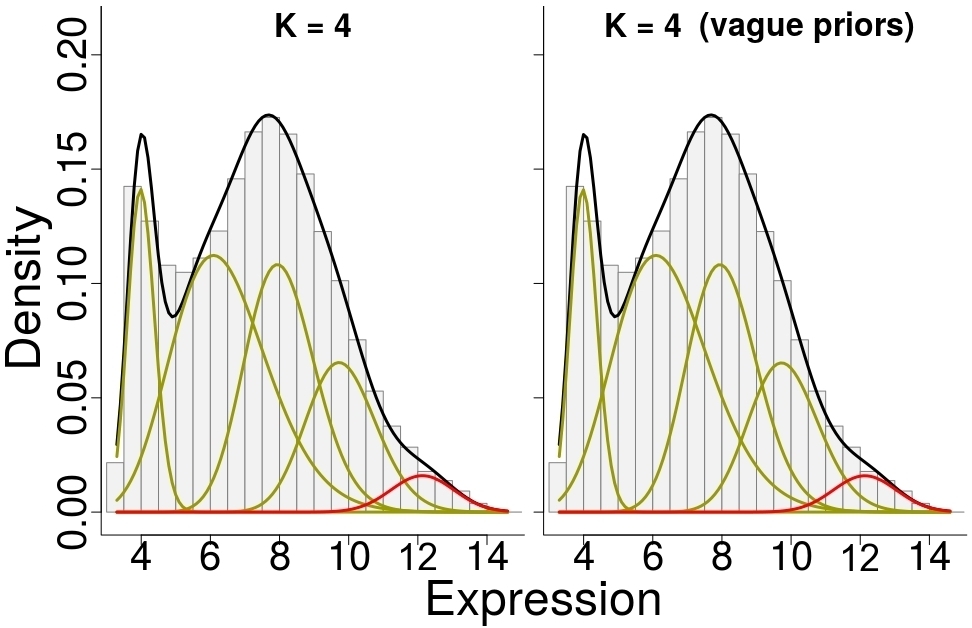}
\vspace{-10pt} \caption{\scriptsize Histogram of all expressions (``Breast 1'' data) overlaid by the estimated mixture density (black curve) and its components (gammas in yellow, normal in red), for the two prior specifications.}
\label{SAP1}
\end{figure}

We also perform a prior sensitive analysis for the parameters of the mixture components, with $K=4$. We consider two different prior specifications, where the standard deviations from one of them are twice the standard deviations from the other one. More specifically, we assume in specification 1: $(\mu, \sigma^2) \sim NIG(15,\;25,\; 2.1,\;1.1)$, $\theta_k \sim IG(4, t_{2k})$ with $t_{2k} = 9$, $18$, $27$, $36$, for $k = 1, 2, 3, 4$, respectively; this assigns means ($3$, $6$, $9$, $12$) and standard deviations ($2.12$, $4.24$, $6.36$, $8.48$) for each of the four gamma components. Note that this configuration is the original one explored in Section \ref{secresult}. In specification 2, we set: $(\mu, \sigma^2) \sim NIG(15,\;100,\; 2.025,\;1.025)$, $\theta_k \sim IG(2.5, t_{2k})$ with $t_{2k} = 4.5$, $9.0$, $13.5$, $18.0$, for $k = 1, 2, 3, 4$, respectively; this provides the same means, but larger standard deviations ($4.24$, $8.48$, $12.72$, $16.96$) for each of the four gamma components. Results are shown in Figure \ref{SAP1} and Table \ref{SAK3}. They are visually the same indicating robustness to the prior specification for these parameters.

\vspace{5pt}
\renewcommand{\thefigure}{E.\arabic{figure}} \setcounter{figure}{0}
\renewcommand{\theequation}{E.\arabic{equation}} \setcounter{equation}{0}
\renewcommand{\thetable}{E.\arabic{table}} \setcounter{table}{0}
{\flushleft \textbf{Appendix E: MCMC diagnostics} }
\vspace{10pt}

\begin{figure}[!h]
\centering
   \includegraphics[scale=0.27]{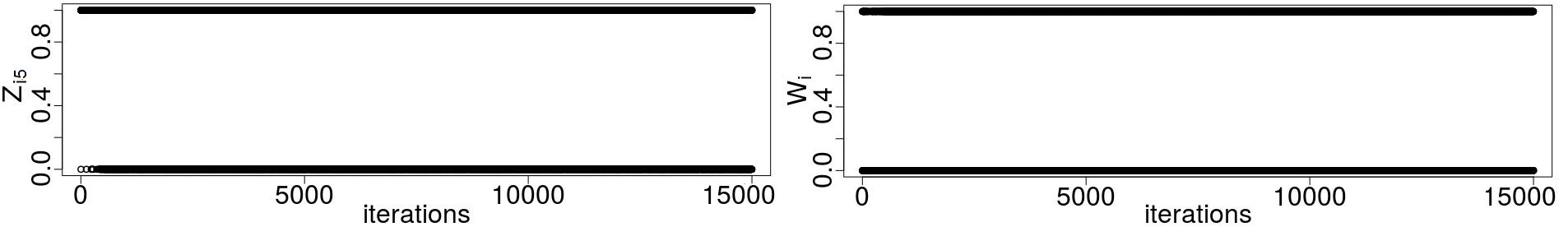}
\vspace{-10pt} \caption{\scriptsize Trace plots of a pair $(Z_{i5},W_i)$ with posterior mean around 0.5 (``Breast 1'' data).}
\label{ZWFC1}
\end{figure}

\begin{figure}[!h]
\centering
   \includegraphics[scale=0.22]{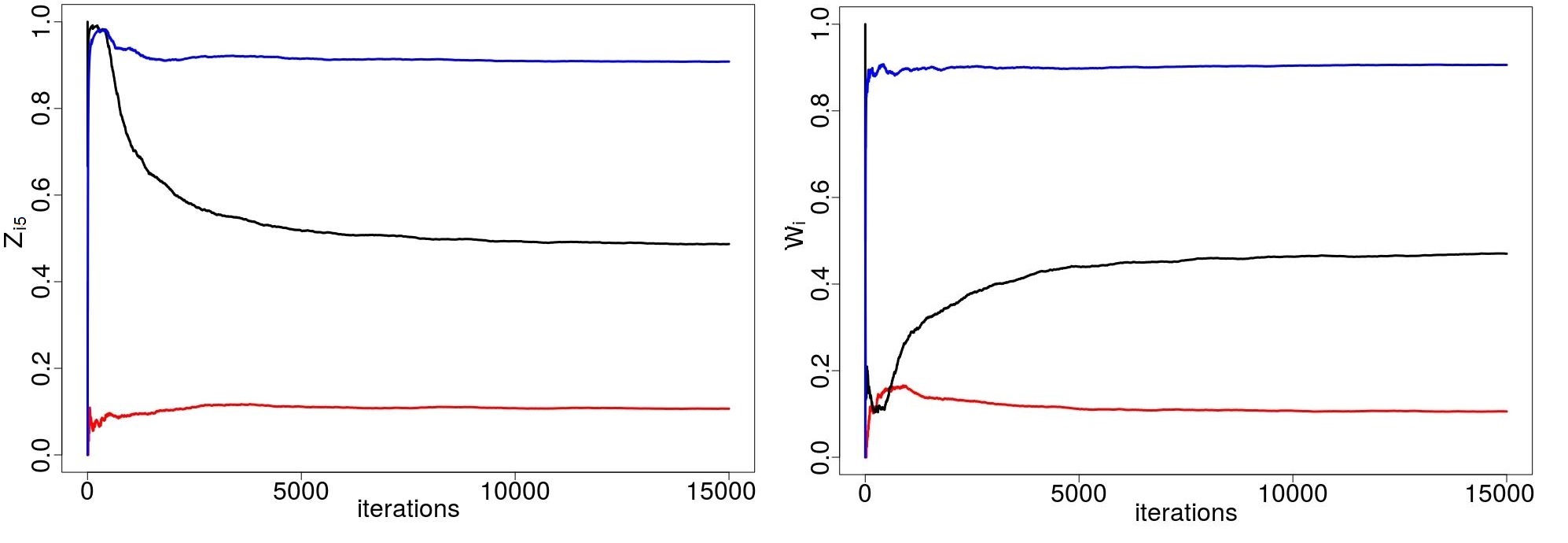}
\vspace{-10pt} \caption{\scriptsize Evolution of the ergodic average of three pairs $(Z_{i5},W_i)$ along the MCMC chain (``Breast 1'' data).}
\label{ZWFC2}
\end{figure}

As we have mentioned in the text, the MCMC we design has good convergence properties. This is due to the chosen blocking scheme and the fact that we can sample directly from the full conditional distribution of $(Z,W)$. Figures \ref{ZWFC1} and \ref{ZWFC2} present some graphs of the MCMC chain for the block $(Z,W)$ to support this point. The trace plots show good mixing properties and the ergodic average trajectories show fast convergence.

\newpage

{\bibliographystyle{Chicago}
\setlength{\bibsep}{0.3pt}
\small \bibliography{references}}

\end{document}